\documentclass[12pt,preprint]{aastex}

\shorttitle{Luminous Stars in N2403 and M81 }
\shortauthors{Humphreys et al.}

\begin{document}

\title{Luminous and Variable Stars in NGC 2403 and M81 \altaffilmark{1}}

\author{Roberta M.. Humphreys\altaffilmark{2}, Sarah Stangl\altaffilmark{2}, Michael S. Gordon\altaffilmark{2}, Kris Davidson\altaffilmark{2}, and Skyler H. Grammer\altaffilmark{2}}

\altaffiltext{1}{Based  on observations  with the Multiple Mirror Telescope, 
a joint facility of the Smithsonian Institution and the University of Arizona 
and on observations obtained with the Large Binocular Telescope (LBT), an international collaboration among institutions in the United
 States, Italy and Germany. LBT Corporation partners are: The University of
 Arizona on behalf of the Arizona university system; Istituto Nazionale di
 Astrofisica, Italy; LBT Beteiligungsgesellschaft, Germany, representing the
 Max-Planck Society, the Astrophysical Institute Potsdam, and Heidelberg
 University; The Ohio State University, and The Research Corporation, on
 behalf of The University of Notre Dame, University of Minnesota and
 University of Virginia.}

\altaffiltext{2}{Minnesota Institute for Astrophysics, 116 Church St SE, University of Minnesota , Minneapolis, MN 55455, roberta@umn.edu}

\begin{abstract}
We present the results of spectroscopy and multi-wavelength photometry of 
luminous and variable star candidates in the nearby spiral galaxies NGC 2403 and M81. We discuss specific classes of stars, the Luminous Blue Variables (LBVs), 
B[e]  supergiants (sgB[e]), and the high luminosity yellow hypergiants. We identify two new LBV 
candidates, and three sgB[e] stars in M81. We also find that some stars previously 
considered LBV candidates are actually field stars. The confirmed
and candidate LBVs and sgB[e] stars  together with the other confirmed members are shown on the HR Diagrams for their respective galaxies. We also present the HR Diagrams for the two ``SN impostors'', V37 (SN2002kg) and V12(SN1954J) in NGC 2403 and the 
stars in their immediate environments. 
\end{abstract}


\section{Introduction} 

This paper is part of a series on the luminous and variable star populations in nearby
resolved galaxies: the giant spiral M101 \citep{Grammer} and the Local Group spirals M31 and 
M33, see Humphreys et al (2017b) and other papers in the series. The primary goal of this work is a  more comprehensive picture of the massive stars that define the upper HR Diagram with special emphasis on an improved census of those that experience high mass loss episodes such as the Luminous Blue Variables (LBVs) \citep{RMH16} and the evolved warm hypergiants \citep{RMH13,Gordon}.  In this paper we present spectroscopy of luminous star candidates in the spiral galaxies NGC 2403 and M81.   

Previous surveys of the stellar populations in these two galaxies began with the classic \citet{ST68} study of NGC 2403.  As part of his survey of the brightest stars in nearby galaxies, Sandage (1984a,b) later presented color-magnitude diagrams for the candidate luminous stars in 
NGC 2403 and M81 based on photometry estimated from photographic plates. The first 
photographic spectra of a few of these stars in NGC 2403 and M81,  previously identified by Sandage but not published, were described by \citet{RMH80}.  
Digital spectra and near-infrared photometry for the red 
supergiant candidates \citep{RMH86} demonstrated that many of them were foreground 
dwarfs.  Several of the blue star candidates were actually H II regions,
especially in M81 \citep{RMH87b}. 
\citet{ZH} later produced  catalogs of multi-color photometry of  
individual stars in NGC 2403 and M81 based on digitized scans of photographic plates. To accurately determine these stars' luminosities and place them on an 
HR Diagram requires confirming spectroscopy.   \citet{Z96} obtained moderate resolution
spectra of the seven visually  brightest blue supergiant candidates in M81, but concluded that most  were  compact H II regions or foreground dwarfs with one possible cluster member.  Subsequently, \citet{Shol98a,Shol98b}  obtained spectra and identified several
emission line objects including LBV candidates in NGC 2403 and M81. Most recently, \citet{Kud} reported the  first quantitative analysis including metallicities, effective temperatures, and luminosities  for 25 early-type supergiants in M81.  

In this paper we present additional blue and red  spectra and  multi-epoch imaging and photometry  
for the  luminous star candidates in NGC 2403 and M81.  In the next section, we describe our target selection, observations, and data reduction.  In $\S3$ we discuss  specific stars such as LBVs and other emission line stars. Multi-wavelength photometry and the spectral energy distributions (SEDs) are presented in $\S4$ for  stars which are candidates for high mass loss. We combine our results with previously published work  and 
present the HR Diagrams for the confirmed members in NGC 2403 and  M81 in the last section.

\section{Data and Observations}

\subsection{Target Selection}

Most of our targets were selected from the \citet{ZH} catalogs for NGC 2403 and M81 based on their apparent magnitude and color. We emphasized those most likely to be 
lumious early-type stars and suspected variables. To select additional candidates, 
we used aperture photometry measured from the Hubble 
Legacy Archive (HLA) Advanced Camera for Survey (ACS) images of  
  NGC 2403 and M81 in the ANGRRR and ANGST programs\footnote{Proposals GO-10182, GO-10579, and GO-10584}.
Known variables from \citet{ST68} and several stars with previously observed 
spectra \citep{RMH87b,Z96,Shol98a,Shol98b} were also included. For the fiber assignment with the MMT/Hectospec,
targets were ranked based on their apparent magnitudes, colors, 
previous spectra suggesting 
that they may be supergiants, and on their variability. Since one of our goals is to identify stars that may  be candidates for high mass loss, we used the 
LBT nearby galaxy survey \citep{CSK,Gerke} to initially identify candidates for 
variability as described in \citet{Grammer}.  
We identified targets as potentially variable if their rms variability is greater than their median photometric error.  The brightest stars, based on their apparent V magnitude with clear indications of variability, received the highest priority, rank 1, for spectroscopy with the Hectospec. The light curves for several of these stars are shown and discussed in \S {3}.  

Following these criteria, we selected 124 stars in NGC 2403 with V magnitudes between 18.0 and 20.1, and 91 in M81 with V magnitudes between 18.5 and 20.1. Eighty-six in NGC 2403 were 
assigned fibers in two separate pointings, fields F1 and F2, and 61 stars in 
M81 were  assigned fibers.  

\subsection{Spectroscopy}  
The spectra were observed with the Hectospec, a multi-object spectrometer mounted on the 
MMT (Fabricant et al. 1998, 2005). The 
Hectospec\footnotemark[1] is a fiber-fed MOS with a 1$^{\circ}$ FOV 
and 300 fibers; each fiber subtends $1.5\arcsec$ on the sky.  
We used the 600 mm$^{-1}$ grating with the blue tilt centered on 4800{\AA} and the red 
tilt centered on 7300{\AA}. The red tilt was chosen to include H$\alpha$ plus the Ca II triplet at $\sim$ 8500{\AA}.  
 The 600 mm$^{-1}$ grating gives  a spectral coverage of $\sim2500${\AA} 
  with 0.54{\AA} pixel$^{-1}$ resolution. The Journal of observations is given in Table 1. 
 
 The NGC 2403 targets were  observed in October and December 2012 in two fields, F1 and 
 F2. The total exposures times in the blue  were 4H and 4H 15M for F1 and F2, respectively 
 and for the red spectra, 3H for F1 and 2.5H for F2. The two pointings overlapped so that 27
 stars 
 were in common yielding total integration times of 8.25H in the blue 
 and 5.5H in the red for these stars. Unfortunately, the M81 observations were plagued by poor observing conditions. Consequently,
 the spectra were acquired over two observing seasons in 2012 and again in 2014. 
 The three best sets of blue spectra were combined to give a total 
 integration time of
 6.75H. Since two of the data sets were separated in time, the spectra provide an opportunity to  
 check for spectroscopic variability before being combined. The red spectra were 
 likewise observed in the two seasons for a  total integration time of 4.5H.

The spectra were reduced using an exported version of the CfA/SAO SPECROAD package for Hectospec data E-SPECROAD\footnotemark[2].  The
   spectra were bias subtracted, flat-fielded, wavelength calibrated, 
   and sky subtracted. The reduced spectra are available at http://etacar.umn.edu/LuminousStars/NGC2403M81/.  

\footnotetext[1]{http://www.cfa.harvard.edu/mmti/hectospec.html}
\footnotetext[2]{External SPECROAD was developed by Juan Cabanela for use on Linux or MacOS X systems outside of CfA. It is available online at http://iparrizar.mnstate.edu.}

\begin{deluxetable}{llccl}
\tablecaption{Journal of Observations\label{tab:one}}
\tabletypesize{\footnotesize}
\tablecolumns{5}
\tablewidth{0pt}
\tablenum{1}
\tablehead{
\colhead{Target}    & 
\colhead{Date}      &
\colhead{Exp. Time} &
\colhead{Grating, Tilt} & 
\colhead{Comment} \\
\colhead{}          &
\colhead{(UT)}      &
\colhead{(minutes)} &
\colhead{}  &   
\colhead{} 
  }
\startdata
NGC2403-F1 Red & 2012 Oct 10 & 180 & 600l, 7200\AA & \\
NGC2403-F2 Red  & 2012 Nov 4 & 150 & 600l, 7200\AA & \\
NGC2403-F1 Blue & 2012 Dec 4 & 120 & 600l, 4800\AA & \\
NGC2403-F2 Blue  & 2012 Dec 4 & 225 & 600l, 4800\AA & \\
NGC2403-F1 Blue  & 2012 Dec 5 & 120 & 600l, 4800\AA & \\
NGC2403-F2 Blue  & 2012 Dec 5 & 30 & 600l, 4800\AA & \\
                 &            &    &              & \\ 
M81 Blue         & 2012 Feb 22 & 150 & 600l, 4800\AA & partly cloudy\\ 
M81 Red          &  2012 Feb 22 &  90  &  600l, 7200\AA & partly cloudy \\
M81 Blue         &  2012 Mar 15 & 120  &  600l, 4800\AA & \\
M81 Blue         &  2014 Feb 20 & 135  & 600l, 4800\AA & clouds, high Z, 
not used\\
M81 Blue         &  2014 Feb 21 & 135  & 600l, 4800\AA & \\
M81 Red          &  2014 Feb 28 & 180  &  600l, 7200\AA & \\
\enddata
\tablenotetext{\,}{`F1' was  centered at 07:36:25.89 +65:38:48.4 and `F2' 
centered at 07:36:23.19 +65:34:54.2}
\end{deluxetable}

\section{Classification of the Stars in NGC 2403 and M81}

Spectral classification of the confirmed members are given in 
Tables 2 and 3 for NGC 2403 and M81. The tables also include 
 positions, 
visual magnitudes, the target source, and comments on the spectrum and 
observations. Non-members or foreground stars plus some likely background QSOs are listed in Table A1 in 
Appendix A. We also include snap-shot images when available in  Appendix B.

In the following subsections we describe specific stars of interest with 
examples of their spectra and light curves for the variables.

\subsection{Emission Line Stars,  Hot Supergiants, and WR Stars}

Although our selection criteria favored  the visually 
brightest stars of spectral types A and F, we identified ten hot or emission 
line stars in NGC 2403 and six in M81.  Three B[e] supergiant candidates in M81 and 
two WN stars in N2403 are described here.  

The B[e] supergiants share many spectral characteristics with LBVs \citep{RMH.4} including prominent Fe II and [Fe II] emission. In a recent  
 spectroscopic survey of emission-line stars, \citet{Aret} designated [O I] 
$\lambda\lambda$6300,6364 emission as one of the characteristics of the sgB[e] 
class. [O I] emission is not observed in confirmed LBVs and can be used 
to separate the two types \citep{RMH.4}.  But these lines are also 
present in the night sky spectrum and in H II regions. For faint stars in N2403 and M81 this can be a problem, with contaminating nebulosity in the aperture 
and residual or poor sky substraction. For that reason, we rely on the velocity of the [O I] lines to identify them with the star, i.e. if they have the same 
velocity as the He I  and the Fe II lines presumably formed in the circumstellar ejecta, although they may  be nebular in origin. 

\begin{figure}[h]  
\figurenum{1}  
\epsscale{1.0}
\plotone{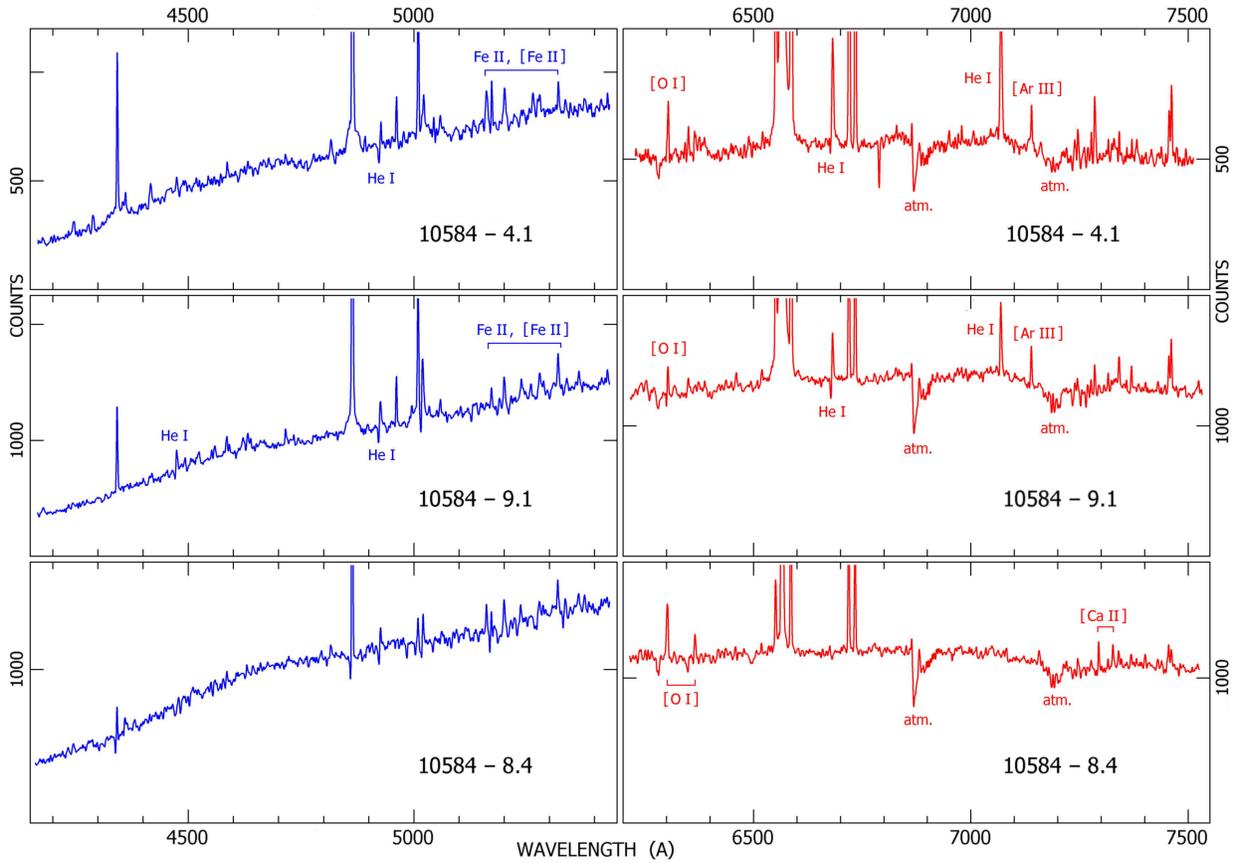}
\caption{Blue and red  spectra of three B[e] supergiants in M81.}
\end{figure}

With this criterion, we identified three  possible sgB[e] stars in M81: 10584-4.1, 10584-8.4, and 10584-9.1. Their blue and red spectra are shown in Figure 1. All three stars have prominent Balmer
emission with P Cyg profiles  and Fe II and Fe[II] lines in the blue plus the  
$\lambda$6300,6363 [O I] lines in the red. He I is also in emission with P Cyg profiles in 
all three stars. Thus they all  show evidence for stellar winds and mass loss. Their  outflow 
velocities,    measured from the absorption minima in their P Cygni profiles relative to the 
emission line peak indicate moderate wind speeds of 170 -- 180 km s$^{-1}$ for 10584-8.4 and 
10584-9.1 and 250 km s$^{-1}$ for 10584-4.1.  These velocities, measured in the same way,  are 
typical of  other sgB[e] stars as well as LBVs \citep{RMH16}.
We note that 10584-4.1 and 10584-9.1 are spectroscopically very similar. Both also have  broad Thomson  scattering wings on  their H$\alpha$ and H$\beta$ emission profiles.  

The spectrum of 10584-8.4 shows  several absorption lines including some  He I lines which, 
together  with other lines such as Mg II, $\lambda$4481 permit us  to estimate a late B/early A   spectral type for this star.  10584-8.4 also has the [Ca II] doublet at $\lambda\lambda$7291,7324 in emission, another characteristic of some of the sgB[e] stars, 
 shared with the warm hypergiants \citep{RMH.4}. The  
Ca II triplet line near $\lambda$8500 can also be seen in emission at the red edge 
of the  spectrum. It shows a split profile which could be due to a bipolar outflow or 
rotation, a characteristic also observed in some sgB[e]'s. The peaks are separated by 108 km s$^{-1}$. 
10584-4.1 and 10584-9.1 are apparently much warmer stars. There are no obvious absorption lines
in either spectrum.  In addition to strong He I, the OI lines at $\lambda$7774 and $\lambda$8446  are also in emission in 10584-4.1.

Light curves for 10584-4.1 and 10584-9.1 are shown  in Figure 2, but  
there is insufficient  data for 10584-8.4.   10584-4.1 shows significant variability over five years by 0.2 to 0.4 mag in the U,V and R bands.  10584-9.1 declined by about 0.5 mag 
from 2011 to 2013.  Thus both stars are variable. 

\begin{figure}[h]
\figurenum{2}
\epsscale{1.2}
\plottwo{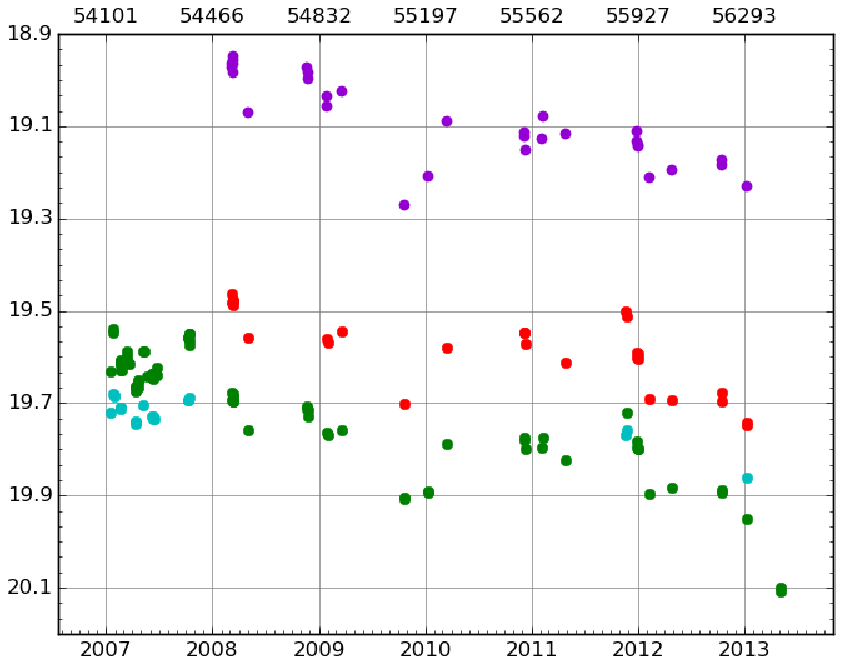}{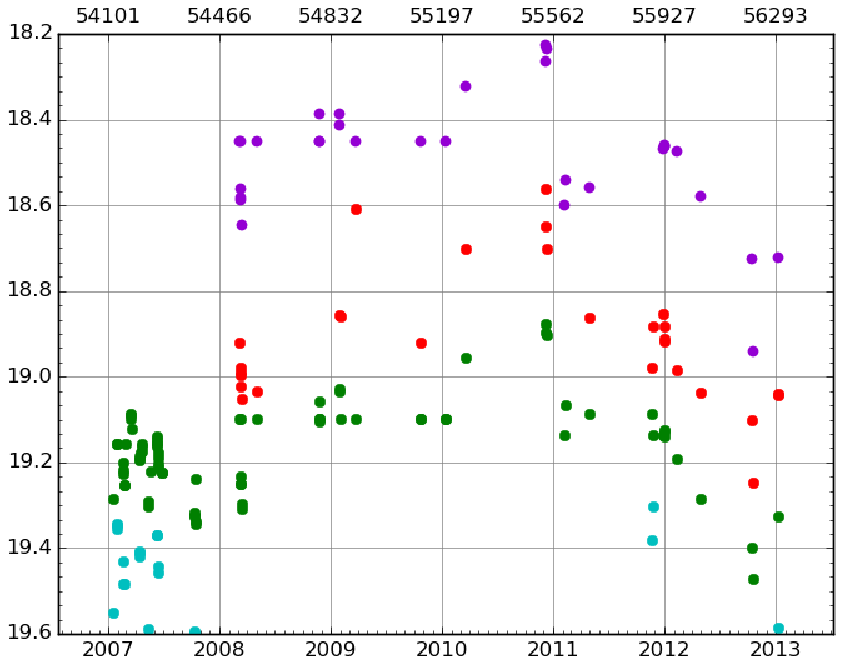}
\caption{Multi-color light curves for the B[e] supergiants 10584-4.1 and 10584-9.1. The {U} band measurements are shown in violet, {B} band in blue, {V} in green and {R} in red.  The formal errors in the LBT/LBC photometry are smaller than the dots.}
\end{figure}

We also indentify two late WN stars in N2403: 10182-pr-9 and ZH 2016. The blue spectra of both  show  prominent broad nitrogen emission features from 4630 to 4700{\AA} and from 5670 to 5700{\AA} with strong H and He I emission lines. He II $\lambda$ 4686 emission is present in 10182-pr-9.

\subsection{Intermediate-Type  Supergiants}

The A- and F-type supergiants are the visually brightest stars. They define the upper luminosity boundary in the HR diagram for evolved post-main sequence massive stars with initial masses typically less than 40--50 M$_{\odot}$ \citep{HD79,HD94}. Many stars that lie near this boundary show 
evidence for high and episodic mass loss in their spectra and spectral energy distributions \citep{RMH13}.  Blue and red spectra of two  luminous intermediate-type supergiants are shown 
in Figure 3.  

ZH 553 in N2403 is a late A-type supergiant of high luminosity. 
It is star IVa28 in previous publications \citep{RMH87a,RMH87b}. Its red spectrum 
shows H$\alpha$  plus the [NII] and [SII] nebular lines in emission. The H$\alpha$ profile 
is asymmetric to the red with  broad wings  
characteristic of Thomson scattering in its wind 
plus P Cygni absorption. The star is thus experiencing  mass loss with a moderately 
slow  outflow velocity of 98 km s$^{-1}$    measured from the P Cyg absorption 
minimum relative to the emission peak. The nebular lines appear to be double peaked. We noticed this in several stars in our similar survey of 
stars in M101 \citep{Grammer}, which in M101 we suspected may  be due to emission from  two sides of large  H II regions or from more than one emission
region along the line of sight. Since ZH 553 does not have  strong [O III] 
nebular emission lines in the blue, the double peaks may rise either from contaminating emission in the aperture plus nebular emission from its circumstellar ejecta, 
or from the two sides of its 
expanding ejecta. The velocity difference between the blue and red components 
average 56 km s$^{-1}$.  There are no other emission lines in the blue or red spectra.   

\begin{figure}[h]  
\figurenum{3}
\epsscale{1.0} 
\plotone{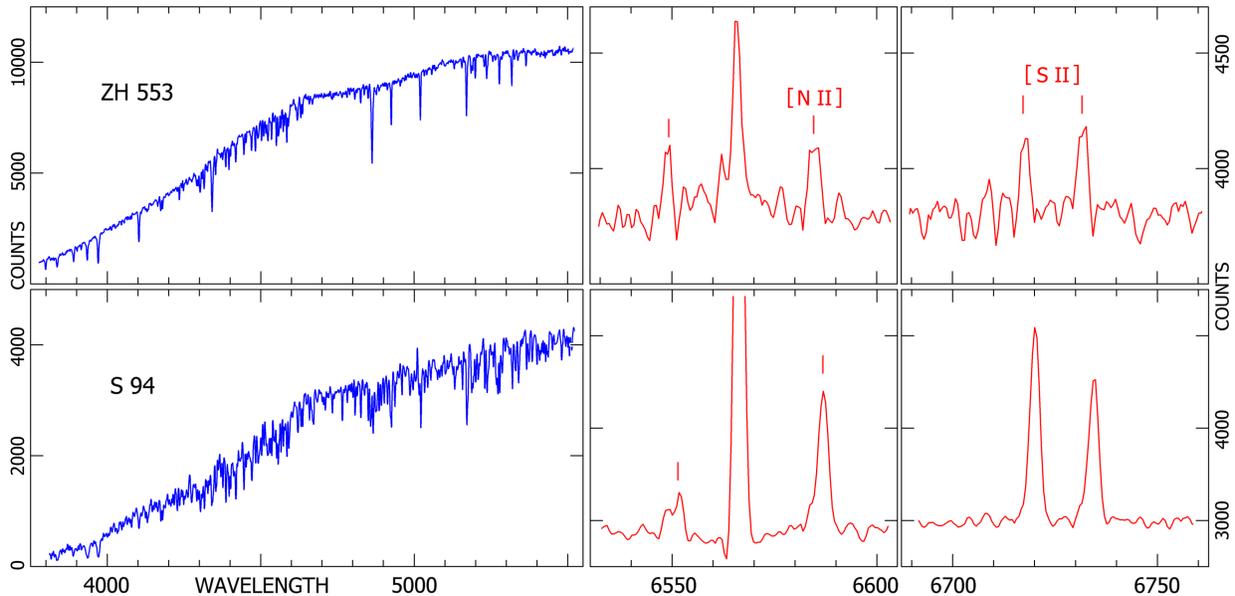}
\caption{Blue and red spectra of two intermediate-type supergiants ZH 553 (A8 Ia) and S94 (F5 Ia). Note the double peaked nebular emission lines in ZH 553, and a probable P Cygni absorption feature in the broad asymmetric  H$\alpha$ emission profile. S94 also has a P Cygni absorption 
feature at H$\alpha$.}  
\end{figure}

S94 was originally listed by \citet{Sandage84a} as one of the brightest 
resolved objects in N2403. It has the spectrum of a high luminosity F-type supergiant. The red spectrum shows H$\alpha$ plus the [NII] and S[II] nebular lines in emission, and  [OIII] $\lambda$5007 is visible in emission in the blue spectrum. H$\alpha$ has a P Cygni absorption feature. The absorption mimimum relative to the peak emission has an expansion velocity of  143 km s$^{-1}$ 
indicating a slow wind and mass loss typical of luminous intermediate type supergiants.   

Both ZH 553 and S94 have marginal variability of 0.1 mag or less.  Their Balmer emission lines with  P Cyg profiles and evidence for mass loss are typical of luminous intermediate-type 
supergiants,  but neither show evidence for circumstellar dust (\S {4.1}).
We don't consider either to be an LBV/S Dor candidate. 

\subsection{Luminous Blue Variables and Candidates}

The \citet{ST68} survey identified several irregular blue variables in N2403. The most famous is V12, also known as SN 1954J, which had a non-terminal giant eruption. Another, V37 also 
received a supernova designation as SN 2002kg due to what was soon recognized as 
a typical LBV/S Dor  high mass loss or maximum light
event. In a recent paper on these two ``impostors'',  we discussed the spectrum and 
light curve of V37 in some detail, and showed that its 
progenitor  was an evolved massive star of $\sim$ 60M$_{\odot}$ \citep{RMH2017}. V12 or SN 1954J survived its giant eruption, and is now obscured by circumstellar dust from that eruption. Our spectral analysis revealed that V12 is actually two stars: a $\sim$ 20,000K star which is the probable progenitor and survivor of the giant outburst,  
plus a G-type supergiant close neighbor or companion. Interestingly we find that the hot star was initially only about 20M$_{\odot}$ and the G supergiant of
slightly lower mass.  The HR Diagram for V12 and V37 and their stellar environments 
is discussed in \S {5}.

Here we include spectra of two additional blue variables in N2403: V38 and V52. Based on the low resolution spectrum of V38 shown in \citet{RMH87b},
we suggested that it was an LBV or LBV candidate. Our higher resolution blue and red spectra and its variability now confirm that designation. Its blue spectrum (Fig.4)  
shows strong nebular and Balmer emission with He I and Fe II emission lines. N II absorption lines are present at $\lambda$5660 - 5680{\AA}. H$\alpha$ has broad wings and a P Cyg 
absorption feature is present at H$\beta$. The [O I] emission lines in its red spectrum at $\lambda\lambda$6300,6363 are also present, at the same velocities as the nebular lines. Thus we suspect that they are 
nebular in origin and the star is not a sgB[e]. Its light curve in Figure 5  shows short-term variability of a few tenths of a magnitude. 

\begin{figure}[h]  
\figurenum{4}
\epsscale{0.6}
\plotone{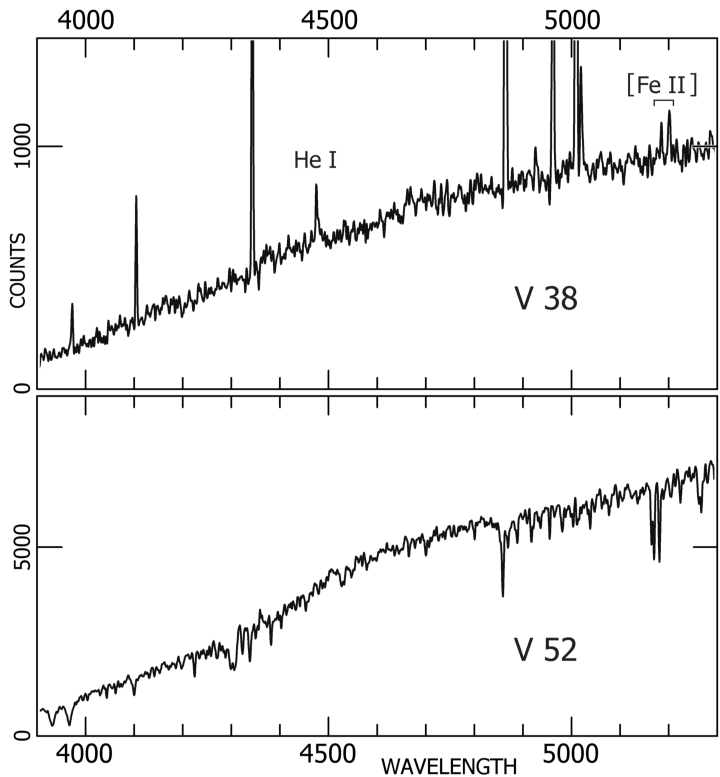}
\caption{Blue spectra of V38 and V52 in N2403. V38 is an LBV while  V52 is a 
foreground F-type dwarf.}
\end{figure}

\begin{figure}[h]  
\figurenum{5}  
\epsscale{0.5}
\plotone{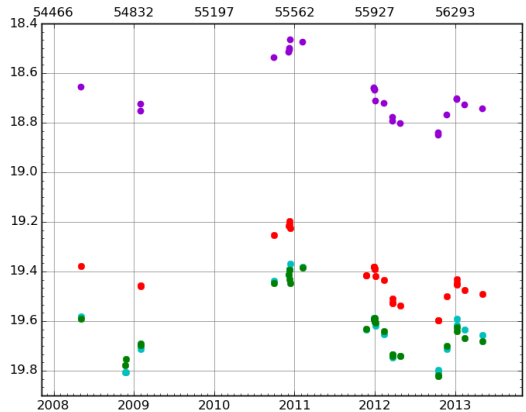}
\caption{The multi-color light curve for N2403-V38. The color code is the same 
as in Fig 2.}
\end{figure}

V52 however is a foreground late F-type dwarf (Fig. 4). Its variability 
reported by \citet{ST68} was marginal, and the LBT survey did not show any 
variability.  

\citet{Sandage84b} identified six irregular blue variables in M81. We observed four 
of them: I1 (ZH 244), I2(ZH 364), I8(ZH 1406), and I3. I3 is a foreground F5 dwarf.  I2 was observed in our earlier study
\citep{RMH87b}, and based on its low resolution spectrum we considered it a 
candidate LBV(LBVc). Our new blue and red spectra (Fig. 6)  reveal a complex spectrum with 
emission lines   of H, He I, and Fe II and [Fe II]. Strong nebular emission lines are  also present. The nebular lines have a somewhat different average velocity of $\approx$ -130 km s$^{-1}$ compared to the He I, H and Fe II emission with velocities of -80 to -100 km s$^{-1}$. The [O I] lines at 6300{\AA} have velocities of -125 km s$^{-1}$ so we assume that they are nebular in origin and are not from the star.  

\begin{figure}[h]  
\figurenum{6}  
\epsscale{1.0}
\plotone{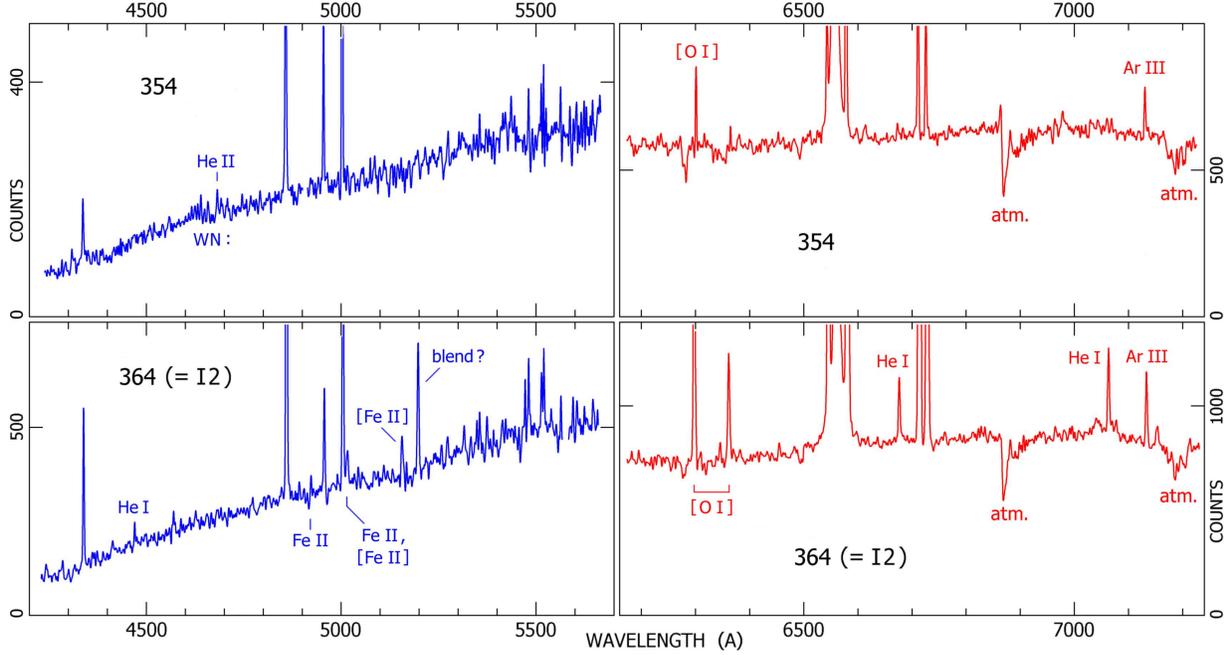}
\caption{Blue and red spectra of two LBV candidates in M81: ZH 354 and I2
(ZH 364).  A small gap at $\sim$4900{\AA} in ZH 354 is due to a flaw in the data.}
\end{figure}

I8 is an F5 supergiant. Its blue spectrum does not show any emission lines. The red spectrum has strong nebular emission but no other emission lines. I8's variability is marginal. Thus, we do not consider it an LBV or candidate. It is worth noting that during their 
high mass loss or dense wind state, LBV/S Dor variables  have absorption line spectra that resemble A or F-type supergiants due to their optically thick cool winds. Thus the ``less luminous''  LBV/S Dor variables \citep{RMH16}, when in ``eruption'',  overlap the position of the normal luminous intermediate temperature supergiants on the HR Diagram, but LBVs in ``eruption'' also have strong H emission with prominent P Cyg profiles and Fe II emission. 

The blue spectrum of I1 has low S/N limiting the accuracy of the classification, but its lines are consistent with a late B or early A-type supergiant. The only emission line is H$\alpha$ with wings which are asymmetric to red. I1 apparently has a stellar wind but there are no  P Cyg features or other emission lines. Its light curve (Figure 7) however shows a pattern of smooth variability over five years. It may be similar to M33C-4640 \citep{RMH16,RMH.4}, a candidate for post-RSG evolution.   

ZH 354 in M81 is another emission-line star that we include here as a candidate LBV (Fig. 6). 
The [O I] $\lambda\lambda$ 6300,6363 lines are present, but their near zero velocities compared
 with -220 to -250 km s$^{-1}$ velocities of the other emission lines confirm that they are 
 residual night sky lines.  ZH 354 also  has features  in common with the Of/WN stars.  
 Its  spectrum shows strong Balmer emission, nitrogen  emission in the $\lambda$ 4600 to 4700{\AA} region  with He II $\lambda$4686 in emission. There is no obvious He I emission in the 
 rather low S/N spectrum.  H$\alpha$ has very broad wings but with no P Cygni absorption.  Its light curve  shows only marginal variability of $\pm$ 0.1 mag over five years.  Many LBVs in their quiescent state have Of/late WN 
 features, so it is possible that ZH 354 is an LBV in quiescence.

The light curves of I2 (ZH 364), I1 (ZH 244) and ZH 354  are shown in Figure 7.

Based on our spectra and the available light curves, V37 and V38 in N2403 are confirmed 
LBV/S Dor variables, and I2(ZH 364) and ZH 354 in M81 should be considered candidate LBVs. 

\begin{figure}[h]  
\figurenum{7}
\epsscale{0.7}
\plotone{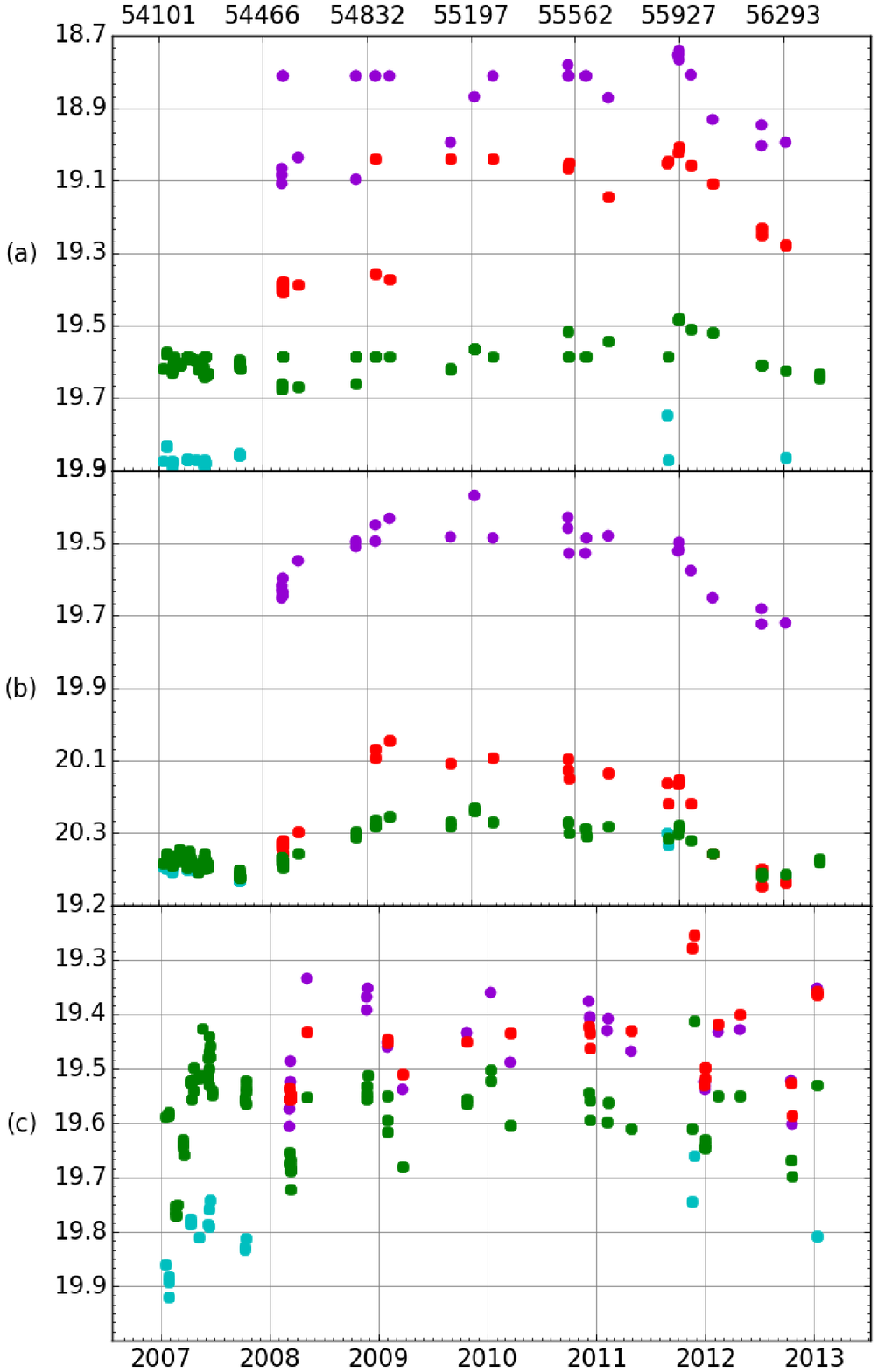}
\caption{The light curves for a. I2(ZH 364), b. I1(ZH 244) and c. ZH 354. The color code is the same as in Figure 2.}
\end{figure}

\subsection{Comparison with other Surveys}

In addition to our previous papers \citep{RMH87a,RMH87b}, Sholukhova et al. (1998a,b) observed 
several of the stars listed by Sandage(1984a,b) plus others from \citet{ZH} and \citet{Z96}. 
Based on their classification of several as LBV candidates, we included them on our program but
found that most are not LBVs.   
As already described above, S94 is a very luminous F-type supergiant but not an LBV candidate while V52 is a foreground dwarf. For convenience and ease of comparison we list the stars in 
common with previous work in Table 4, together with our classifications in 
this paper.

\section{Interstellar Extinction, the Spectral Energy Distributions and Circumstellar Dust}

The lack of a uniform dataset for the visual photometry and limited  
coverage in the near-infrared in  N2403 and M81 complicates
a comprehensive survey of the spectral energy distributions for most of the
confirmed members. 

The visual-wavelength photmetry comes from three sources, the \citet{ZH}
photographic survey, the GO fields in the ANGRR and ANGST programs with HST/ACS and the 
LBT/LBC survey. No one survey includes all of the stars in Tables 2 and 3. The HST fields of course have the highest spatial resolution but usually  
include only two colors, visual and blue.  The LBC dataset includes UBVR\footnote{The UBV magnitudes are on the Johnson system but R is Cousins-Kron.} 
magnitudes, but is seeing limited.    
 The formal errors listed for the stars in the LBT/LBC survey are quite small, typically 0.03 -- 0.02 mag and less, and are smaller than the dots on the 
light curves shown in Figures 2, 5 and 7. The majority of stars in Tables 2 and 3 are in this set.  

In addition, we used the  DOLPHOT package for WFC3/IR to measure near-infrared 
VEGA magnitudes at 1.1 and 1.6 microns from GO11719 and GO 13477 for N2403 and from  GO12531 and GO11731 for M81.
Due to the limited spatial coverage, only 8 confirmed members  in N2403, and 9 
in M81  
have measured near-infrared magnitudes.  
We also measured mid-infrared magnitudes from the {\it Spitzer} IRAC surveys. 
We used the median mosaic images in all four 
IRAC bands from the {\it Spitzer} archive. The MOPEX/APEX package was then used 
to measure point response fitting photometry with the detection limit set at 
three sigma. Many of the stars were too faint to be detected, and the 
photometry is further complicated by the high backgrounds and
complex extended regions where they are found. Therefore  each image was 
inspected individually. 

We list all of the available photometry for the confirmed members in Table 5, with 
the exception 
that the photographic magnitudes are listed only if no other source is available.

To determine whether these stars have excess free-free emission from their stellar
winds and/or circumstellar dust, as well as their intrinsic luminosities, we must first 
correct their SEDs for interstellar extinction. For stars with multi-wavelength
visual photometry and spectral types, we adopt the \citet{Cardelli} extinction curve with R = 3.2 and follow the standard procedure and estimate the 
reddening E(B-V) and visual extinction A$_{v}$ from their observed colors and spectral types. However,   broadband colors cannot be safely used for stars with strong emission lines. In our previous 
work on M31 and M33, we adopted the mean extinction from nearby stars  and from the H I column density. However, extensive catalogs of resolved stars in the fields of  N2403 and M81 
do not yet exist and the neutral hydrogen maps have much lower spatial resolution at 
their larger distances. In our detailed study of V37 and V12 in N2403 \citep{RMH2017}, 
we determined visual extinctions of 0.54 mag and 0.9 mag, respectively from the stars in 
their near environments. In this work, we  find a  mean extinction for the confirmed
supergiants of 0.47 mag. We have a similar situation in M81. \citet{Kud} found a mean color 
excess of 0.26 mag or A$_{v}$ of 0.9 mag from their quantitative analysis of 25 early-type 
stars in M81. Our mean extinction for the confirmed supergiants in M81 is a very 
similar 0.86 mag. Therefore, in this study we adopt total  visual extinctions (A$_{v}$) 
of 0.5 and 0.9 magnitudes respectively in N2403 and M81, for 
the emission line stars, the LBVs and sgB[e]'s, and for those stars which lack complete 
photometry. In Table 6, we summarize the results for the confirmed stellar members with 
adopted distance moduli of 27.5 mag for N2403  and 27.8 for  M81 from Cepheids \citep{Freedman2001}  to derive their corresponding absolute visual magnitudes.  

\subsection{ The Spectral Energy Distributions (SEDs)} 

Despite the limitations of the multi-wavelength photometry and especially the lack of infrared data for many of the stars, we show a selected sample of SEDs in Figure 8,  
specifically of stars of interest such as the LBV candidates, the sgB[e]'s and 
and others with possible circumstellar dust. 

The SEDs for the two LBV candidates in M81 are shown in the top panel. Since these are 
strong emission line stars, their photomery is corrected for interstellar 
extinction using the mean A$_{v}$.  
ZH 364(I2) has a near-infrared excess which we attribute  to free-free
emission.  Its spectrum shows a strong H$\alpha$  emission line with broad wings. We also note the raised photometric points in its SED in the R-band due to H$\alpha$ and in the U-band possibly due to continuum emission. Thus its near-infrared excess is most likely due to free-free emission and not warm dust. We show  Planck curve fits to their corrected 
broadband data to estimate the temperature shown and integrated to derive a
luminosity (M$_{Bol}$) in Table 6.  However, the temperature for ZH 354 from the fit to the 
LBT/LBC photometry is inconsistent with the much higher temperature implied by its emission lines such as He II in its spectrum (Figure 6). It is possible that the extinction correction is much larger than 
the adopted mean, or the LBC photometry is identified with the wrong star.  Its luminosity derived from the SED is not used for this reason.   

The SEDs for two  sgB[e] stars with infrared data are shown in the middle 
panel. 
Their mid-infrared fluxes 
demonstrate the presence of significant circumstellar dust as found 
for many sgB[e]'s in other galaxies \citep{Kraus,RMH.4}. Although the optical 
photometry for 10584-8.4 in M81 is limited, its SED exhibits a large circumstellar excess due to dust plus extensive circumstellar gas revealed by the [Ca II] and Ca II emission lines in its red spectrum, \S {3.1}. The mid-infrared fluxes may seem high or elevated with respect to the visual photometry, however this strong infrared signature for 
circumstellar dust is not uncommon for sgB[e]'s \citep{RMH.4,RMH.5}. 
Although the  absorption lines in its blue spectrum suggest a late B/early A classification, the Planck fit to the non-uniform  optical photometry yields a temperature of 
21,700 K. We consider this result doubtful, though, because the fit is  dominated by the 
uncertain U band point from the LBT/LBC imaging.  Using only the HST magnitudes, the best fit yields 10,900 K. 
10584-9.1 is a much hotter star as indicated both by is spectrum and SED. 
The Planck fit to the optical photometry suggests a temperature of 18,000 K. 
Its mid-infrared excess may be a  combination of free-free emission  
and dust. 

\begin{figure}[h]  
\figurenum{8}
\epsscale{0.9}
\plotone{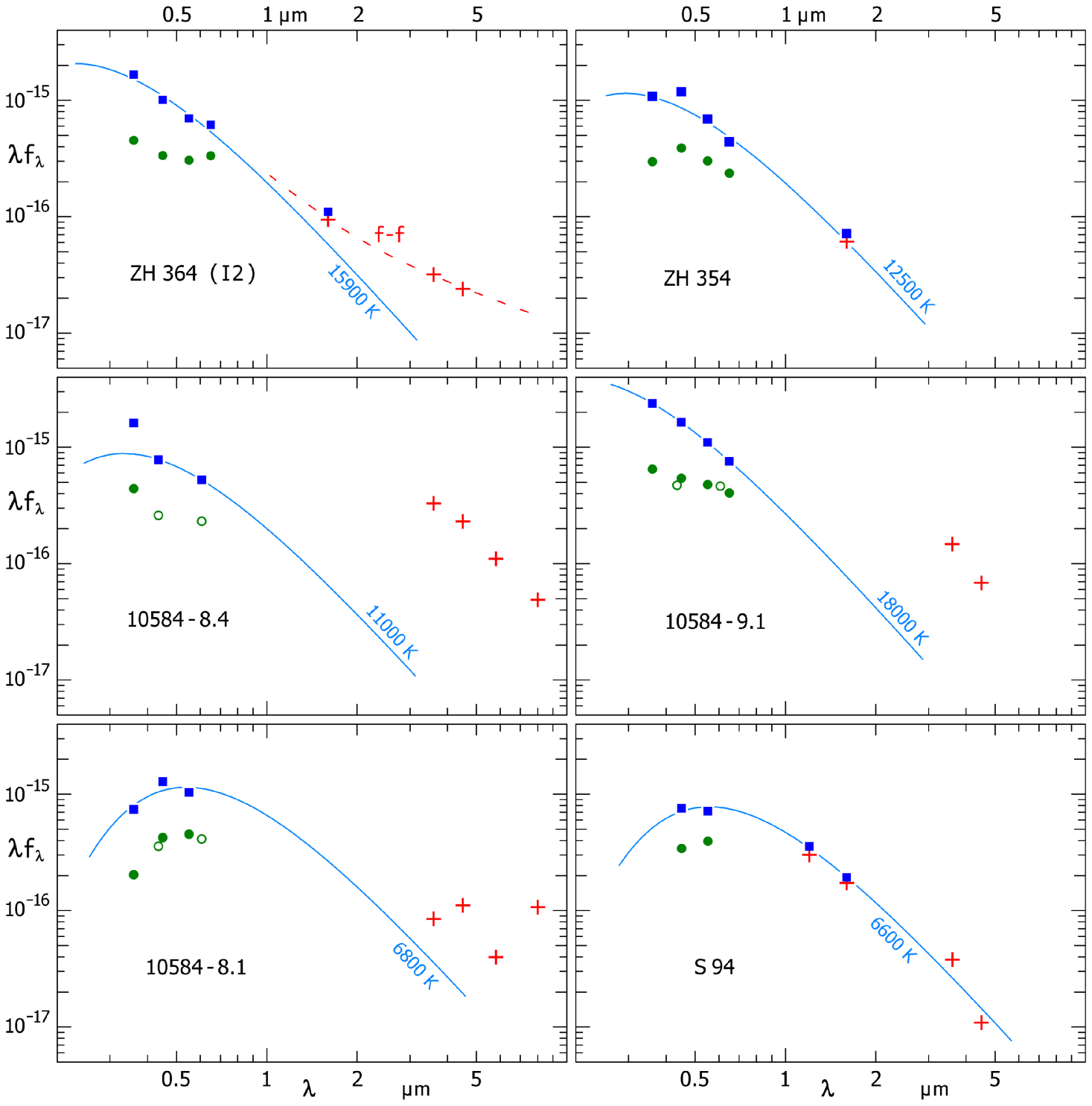}
\caption{The SEDs for six luminous stars to illustrate the presence or lack of circumstellar dust. The green dots are the observed LBT/LBC  visual-wavelength 
magnitudes 
and the blue boxes show the same photometry corrected for interstellar extinction (Table 6). The open circles show the magnitudes from the HST/ACS images when 
available. The red crosses are the near-IR magnitdes from the HST/WFC3 images and from Spitzer/IRAC. The statistical one sigma errors are smaller than the dots in these log-log plots. }
\end{figure}

We also show the SEDs for two luminous intermediate-type supergiants 
in the bottom panel. 10584-8.1 may have circumstellar dust although its spectrum did not show
any stellar wind emission lines. Planck curve fits to their optical 
photometry are shown.

The temperatures and derived luminosities estimated in this way are used to place the emission line
stars on the HR Diagram discussed in the next section.  We note as usual, that Planck
curves are only rudimentary approximations. 

\section{The HR Diagrams} 

The HR Diagrams for the confirmed stellar members in NGC 2403 and M81 from Table 6 are shown in Figures 9  and 10 respectively.  The  bolometric luminosities and temperatures in Table 6 are adopted  from the calibrations by \citet{Flower} for the supergiants and \citet{Martins} for the O-type stars and from the Planck fits to the SEDs for the emission line stars.  We  added the late B- and early A-type supergiants in M81 from \citet{Kud}. The temperatures are derived from the data in their Table 3 and their derived 
luminosiites are corrected to our adopted distance modulus\footnote{\citet{Kud} derived a distance modulus of 27.7 mag for M81, while we have adopted 27.8 mag based on 
Cepheids. A small difference.}.

\begin{figure}[h]  
\figurenum{9}
\epsscale{0.8}
\plotone{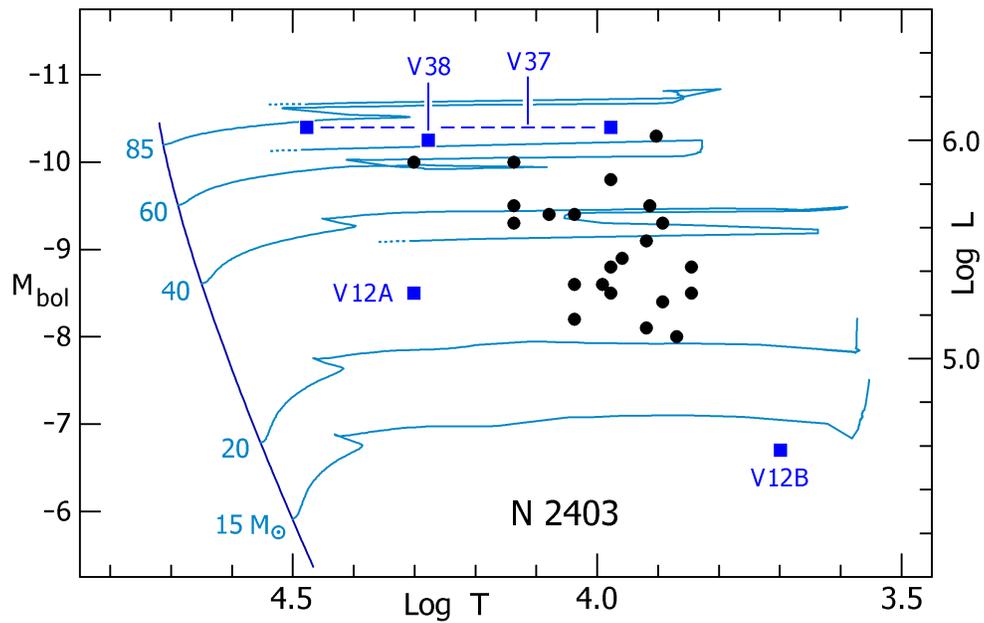}
\caption{The HR Diagram for NGC 2403 stars presented in this paper. 
The positions of the confirmed LBVs V37 and V38 are identified. V37's position 
is shown  during its high mass loss state in 2002 and in quiescence. The surviving star from V12's (SN 1954J) giant eruption and its less luminous 
cooler companion are plotted as V12A and V12B. The evolutionary tracks with mass loss are from \citet{Ekstrom} without rotation.}
\end{figure}

\begin{figure}[h]  
\figurenum{10}
\epsscale{0.8}
\plotone{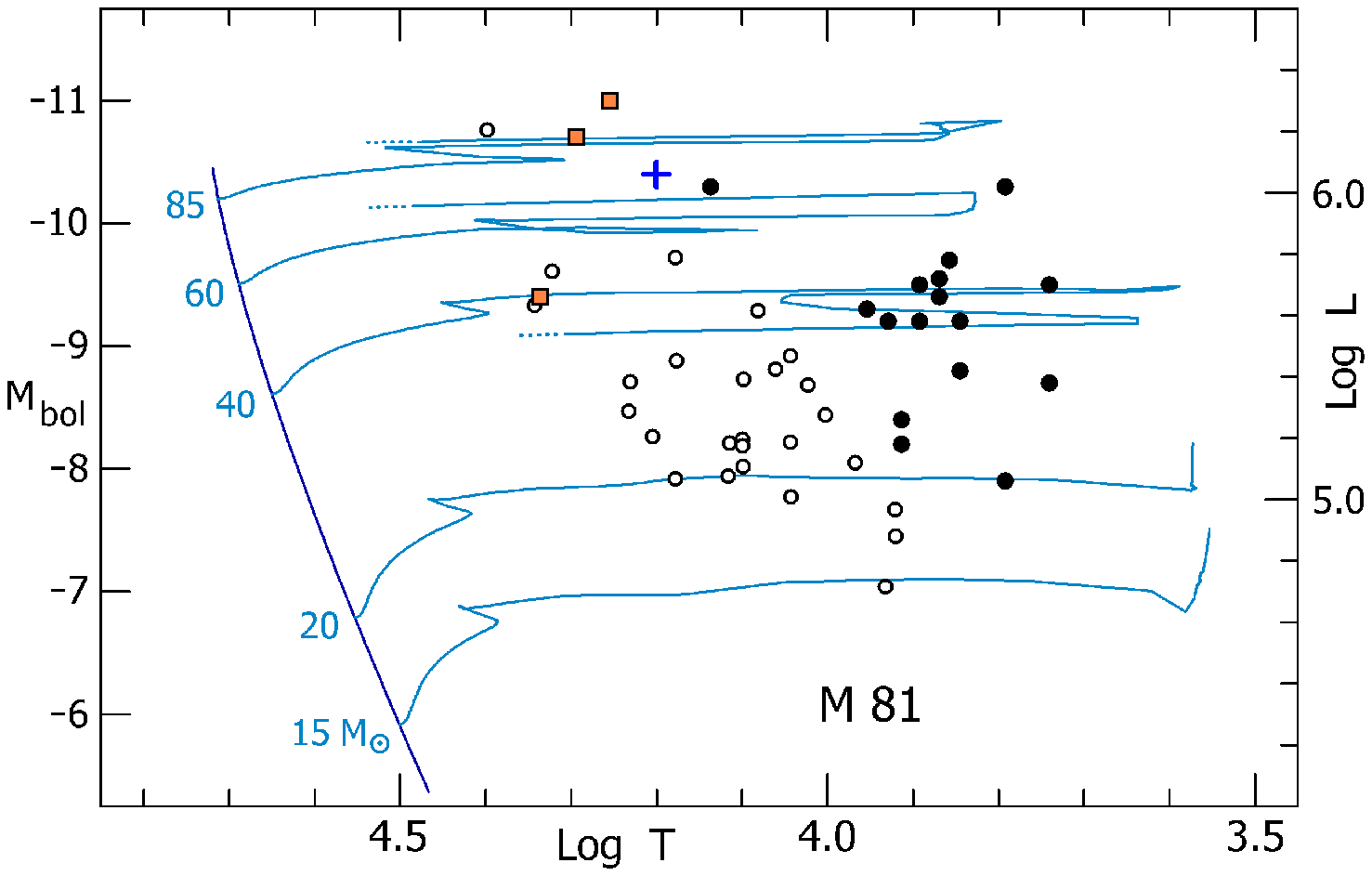}
\caption{The HR Diagram for M81. The stars  presented this paper are plotted as filled circles and those from \citet{Kud} as open circles. The position of the 
candidate LBV I2 (ZH 364) is shown as a blue cross and the three B[e] supergiants as orange squares. The LBV candidate ZH354 is not shown. Its available photometry is not consistent with its spectrum.}
\end{figure}

The uncertainty in the placement of the individual stars on the HR Diagrams will vary from star to star. A major source is the adopted B-V colors and the interstellar extinction. The range in the observed B-V colors (Table 6) is
$\approx \pm$ 0.1 mag which will translate to an uncertainty up to 0.3 mag in 
M$_{V}$.  For example, 
two stars, 10182-pr-6 in NGC 2403 and 10584-13-3 in M81, lie just to right of the upper luminosity boundary or Humphreys-Davidson limit shown in Figure 11. While this could be real and indicative of their evolved 
state, both stars have notably high values of total interstellar extinction (A${_{V}}$), and as was noted in Table 2,  10182-pr-6  may be more than one star  based on its spectral features. 

The LBV V37 (SN 2002kg) and the giant eruption/SN impostor V12 (SN 1954J) both in NGC 2403 are the only two
stars in our survey with available data for their neighboring stars \citep{vandyk,RMH2017}. For that reason we show a
separate HRD in Figure 11 for these two stars and their stellar environments. We used the 
two-color diagram for the stars near V37 \citep{vandyk} plus the Q-method to estimate their intrinsic  {B-V} colors, corresponding 
spectral types, and interstellar extinction from which we drived their visual and absolute bolometric 
magnitudes to place them on  the HRD shown here. 

\begin{figure}[h]  
\figurenum{11}  
\epsscale{0.8}
\plotone{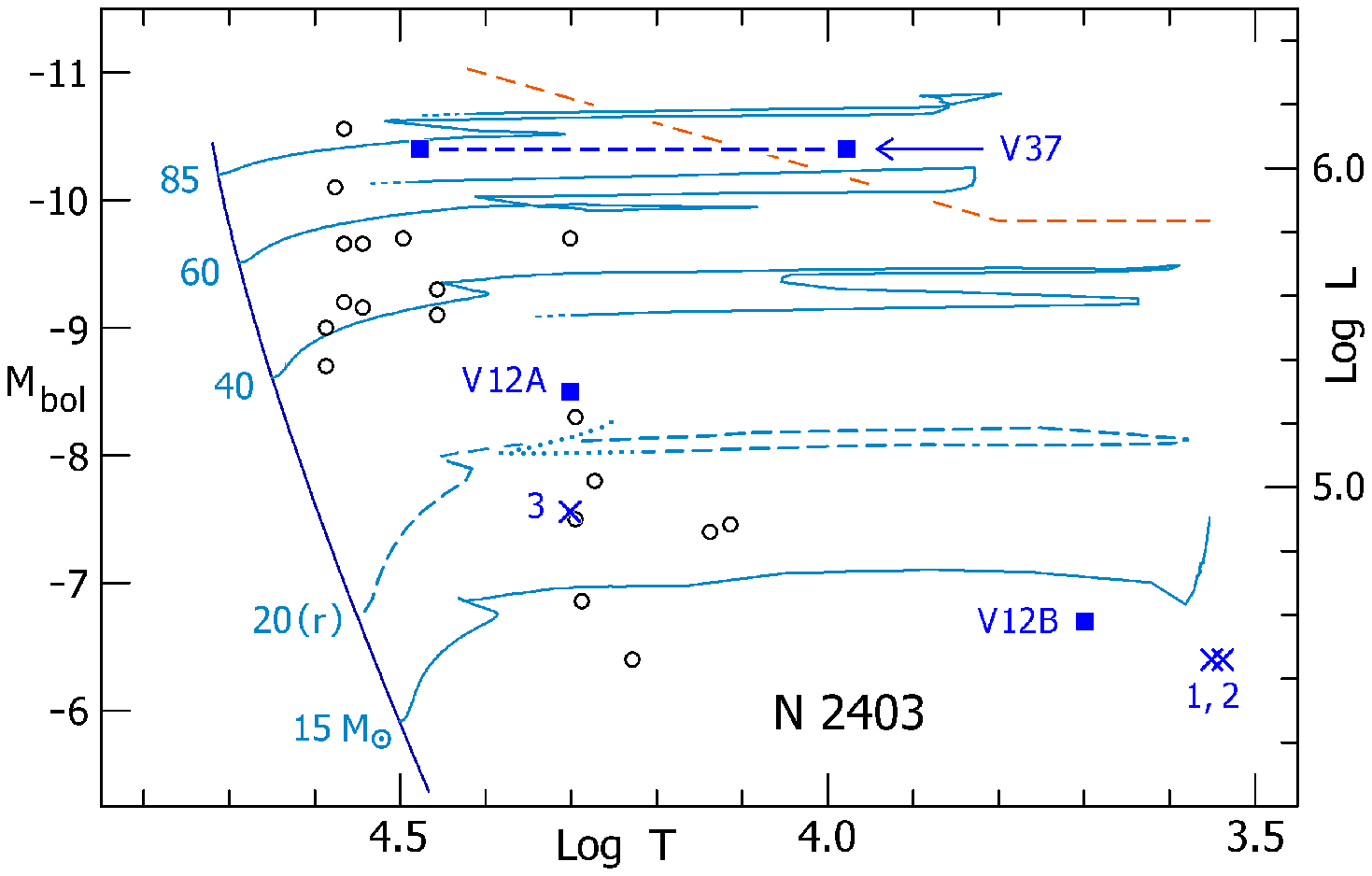}
\caption{An HR Diagram illustrating the stellar population and nearby stars in the near environments of V37, plotted as open circles, and for the giant eruption V12. Its neighbors within $\sim$ 1 arcsec, stars 1,2, and 3, are shown as blue crosses.  V12 itself is not resolved in the HST images but its spectrum shows 
it is two stars; the hot star V12A, the likely survivor, has a cooler less luminous neighbor V12B. In this HRD we show the 20M$_{\odot}$ track with mass loss 
and rotation \citep{Ekstrom}  to illustrate the likely post-red supergiant evolutionary state for V12A. The dashed red line is the upper luminosity boundary \citep{HD79,HD94}} 
\end{figure}

V37 is associated with other reddened hot stars and is one of the most massive in its
environment.  It is not known if V12 is a physical pair but its companion is a G type supergiant, and 
its nearby neighbors include a hot supergiant and  two red supergiants. Thus both are closely associated with other evolved stars. Neither is isolated as has been suggested for some LBVs \citep{Smith2015}.   
Based on its temperature and luminosity, V12(A) will lie just below the LBV instability strip on the HR Diagram \citep{RMH16}. Like the ``less luminous'' LBVs, V12 was very likely a post-red supergiant, and having shed a lot of mass it was close to its Eddington limit at the time of its giant eruption. On Figure 11 we show the evolutionary track for a 20 M${_{\odot}}$ star with mass loss and rotation \citep{Ekstrom} that  illustrates probable  post-RSG evolution. Note the short transit back to cooler temperatures at the end of the track near the likely position of V12A's progenitor, an ideal state for a highly evolved star to experience an eruption.

\acknowledgements
Research by R. Humphreys on massive stars was supported by  
the National Science Foundation grant AST-1109394. We thank Schuyler Van Dyk for sharing the optical photometry for the stars in the near environment of V37(SN 2002kg) used for his Figure 9 in \citet{vandyk}.   

{\it Facilities:} \facility{MMT/Hectospec, LBT/LBC, HST/ACS, HST/WFC3}

\appendix

\section{Foreground Stars}


\section{Snaphot images}
Snaphot images of 12 of the confirmed members in NGC 2403 and 18 in M81 with 
images on  the {\it HST/HLA/ACS} frames.  Each image is 10{\arcsec} on a side and the star is marked.






\begin{deluxetable}{llccccl}
\tabletypesize{\scriptsize}
\rotate
\tablenum{2}
\tablecolumns{7}
\tablecaption{Members of NGC 2403}
\tablewidth{0pc}
\label{tab:n2403_members}
\tablehead{\colhead{Star ID} & \colhead{Position${J2000}$}
  & \colhead{Sp. Type} & \colhead{$V$}  & \colhead{Source\tablenotemark{a}} &  \colhead{Variability} & \colhead{Comments}}
\startdata  
ZH 585 (F1+F2) & 7:35:37.75 65:35:33.77 & F2 I & 19.43  & 1 & \nodata & H$\alpha$ em, neb em\\
ZH 553 (F1+F2) & 7:36:10.10 65:33:30.91 & A8 I    & \nodata  & 1 & no & see text \\
ZH 335 (F1+F2) & 7:36:12.02 65:32:44.39 & F0 I & 19.19  & 1 & no & neb em\\  
ZH 2387 (F1+F2)& 7:36:15.63 65:40:43.15 & H II & 20.2 &  1 & \nodata & He I em\\
ZH 2352 (F1) & 7:36:16.59 65:37:34.58 & H II & 19.65  & 1 & \nodata &  \\ 
ZH 1593 (F2) & 7:36:16.79 65:37:25.45 & H II &  19.28  & 1 & \nodata & strong neb em, He I em \\ 
ZH 755 (F1+F2) & 7:36:19.67 65:39:3.09  & H II & 20.15  & 1 & \nodata & He I, N II $\lambda$5670 em\\
ZH 2341 (F1) & 7:36:20.06 65:37:29.73 & H II & 18.42 & 1 & \nodata & He I em \\ 
ZH 2072 (F1) & 7:36:23.69 65:36:19.58 & H II & 20.13   & 1 & \nodata & He I em\\
ZH 2331 (F2) & 7:36:24.55 65:37:57.83 & H II & 19.94   & 1 & \nodata & strong neb em, He I em\\ 
ZH 2328 (F1) & 7:36:25.14 65:37:57.90 & hot star & 20.04 & 1 & \nodata & neb em, He I em, H$\alpha$ broad wings \\ 
ZH 1521 (F1)(10579-x1-7) & 7:36:25.92 65:35:31.23  & F2 I & 18.59 &  1 & \nodata &  neb em\\
ZH 2306 (F2) & 7:36:37.50 65:37:54.47  & H II & 18.28  & 1 & \nodata & strong neb em, He I em\\ 
ZH 533 (F1+F2) & 7:36:37.57 65:33:33.47 & B5: I   & 19.36 & 1 & no & neb em, He I abs, $\lambda$7774 abs \\
ZH 2313 (F1) & 7:36:39.21 65:39:33.89 & A0-A2  & 19.78 & 1 & \nodata & neb em, H$\alpha$ wings \\ 
ZH 2562 (F1) & 7:36:44.36 65:39:11.25 & B I  & 19.57   & 1 &  \nodata & neb em, He I abs, N II $\lambda$ 5670 em \\ 
ZH 2022 (F1) & 7:36:44.73 65:33:25.87 & B8 I & 19.95  & 1 & \nodata & neb em, H$\beta$ P Cyg + wings, H$\alpha$ broad wings\\
10182-pr-9 (F2) & 7:36:45.49 65:37:0.83 & WN: & \nodata & 2 &  \nodata & [N II] em, H, He I em, H$\alpha$ wings \\ 
ZH 2016 (F2) &  7:36:47.84 65:33:26.08 & WN: & 18.47  & 1 & \nodata & strong neb em, N II, He I em\\ 
10579-x1-3 (F1) & 7:36:48.56 65:36:45.50 & neb em & 20.3  & 3 & \nodata  & H em. superposed on abs.\\ 
10182-pr-1 (F1) & 7:36:48.79 65:35:52.74 & A5-A8 I & \nodata  & 2 &  \nodata & strong neb em \\
ZH 946 (F2) & 7:36:50.63 65:38:49.10 &  B5 I: & 19.74  & 1 & \nodata  & neb em, Horiz Br? \\  
ZH 947 (F1) & 7:36:51.44 65:39:0.47 &  B5 I & 19.06   & 1 & \nodata & neb em, He I abs\\ 
10182-pr-2 (F1) & 7:36:56.20 65:36:42.04 & F5 I & 18.83  & 2 & var & neb em, 
H$\alpha$ P Cyg \\ 
ZH 2554 (F1+F2) & 7:36:58.28 65:41:5.59 & H II & 20.2  & 1 & \nodata & He I em\\
10182-pr-6 (F1) & 7:36:59.12 65:35:9.95 & A8-F0 I & 18.79 & 2 & \nodata & strong neb em, He I $\lambda$6678,7065 P Cyg, Ca II abs, probable blend \\
10182-pr-15 (F2) & 7:36:59.12 65:35:17.97 & A0-A2 I & 19.52 & 2 &  \nodata & strong neb em, He I abs, Mg II abs\\ 
10182-pr-16 (F2) & 7:37:01.34 65:34:26.06 &  B8 I: &  19.35  & 2 & \nodata & He I abs\\ 
ZH-729 (F2) & 7:37:01.62 65:37:31.95 & B8 I & 19.21  & 1 & \nodata & neb em, H em in abs core, He I abs \\ 
V37 (F1) & 7:37:01.83 65:34:29.3 & LBV & 20.6  & 4 & var & see \citet{RMH2017} \\
ZH 2248 (F2)(10402-7) & 7:37:6.56 65:33:54.21 &  A0 I & 19.71  & 1 & var & neb em\\ 
V38 (F1) & 7:37:10.6  65:33:10   & LBV & 19.4 & 4 & var & neb em, see text \\ 
ZH 931 (F1) & 7:37:10.77 65:39:41.59  &  A5-A8 I & 19.95 & 1 & \nodata & H$\alpha$ core em \\ 
ZH 1938 (F1) & 7:37:12.02 65:32:1.15 & A2 I & 19.36  & 1 & var &  neb em \\
S94 (F1) & 7:37:12.79 65:36:12.68 & F5 I & 18.78 & 5 &  var: & see text \\ 
ZH 924 (F1) & 7:37:15.47 65:38:38.04 & B5-B8 I & 19.3  & 1 & \nodata & neb em, He I abs\\ 
ZH 1483 (F2) & 7:37:15.76 65:32:02.11 & H II &  18.54  & 1 & \nodata & strong neb em, He I em \\ 
ZH 2212 (F2) & 7:37:21.07 65:33:5.86 & H II &  19.9 & 1 & \nodata & neb em, He I em\\ 
ZH 912 (F1) & 7:37:32.86 65:38:59.49 & A0-2 I & 18.41  & 1 & \nodata& neb em, He I $\lambda$6678 abs, $\lambda$7774 abs, A8 Ia\citep{RMH87b} \\ 
ZH 884 (F1+F2) & 7:37:48.92 65:35:37.79 & F0 I & 18.78  & 1 & \nodata & neb em, double H$\alpha$ \\ 
 
\enddata
\tablenotetext{a}{Primary Sources for targets: 1) \citet{ZH}, 2) GO-10182, 3) GO-10579), 4) \citet{ST68}, 5) \citet{Sandage84a} } 
\end{deluxetable}

\begin{deluxetable}{llccccl}
\tabletypesize{\scriptsize}
\rotate
\tablenum{3}
\tablecolumns{7}
\tablecaption{Members of M81}
\tablewidth{0pc}
\label{tab:m81_members}
\tablehead{\colhead{Star ID} & \colhead{Position${J2000}$}
  & \colhead{Sp. Type} & \colhead{$V$} & \colhead{Source\tablenotemark{a}} &  \colhead{Variability} & \colhead{Comments}}
\startdata  
ZH 501 & 9:54:34.86 69:05:53.72 & H II & 18.19 & 1  & \nodata & strong neb em, He I em \\
10584-11-3 & 9:54:41.4514 69:04:08.81 & H II & 20.08 &  2 & \nodata & strong neb em, He II em, hot star blend? \\ 
10584-11-1 & 9:54:42.48 69:2:57.04 & A5 I & 19.38 & 2 & var & H em \\ 
10584-11-2 & 9:54:42.57 69:03:38.08 & H II & 19.81 &  2 & \nodata & strong neb em, He I em \\ 
ZH 679  & 9:54:45.40 69:9:26.42 & A8 I & 19.7  & 1  & \nodata  & H em \\ 
10584-8-4 & 9:54:50.03 69:6:55.47 & sgB[e] & \nodata  & 2  & \nodata & H em P Cyg,He I, Fe II, [Ca II] em, see text \\
10584-4-1 & 9:54:54.05 69:10:23.00 & sgB[e] & 19.68  & 2 &  var &  H em, P Cyg, He I, Fe II em, see text \\ 
ZH 372(10584-15-1) & 9:54:56.92 69:01:3.67 & F5 I & 19.64  & 1,2 & \nodata &  \\  
ZH 1434 & 9:55:00.79 69:13:4.32 & H II & 20.12 & 1 & \nodata & strong neb em, He I em \\  
10584-8-1 & 9:55:01.39 69:07:06.02 & F0 I & 19.16  & 2 & var: & \\ 
10584-8-2 & 9:55:09.01 69:07:08.27 & F0 I & 19.46  & 2 & \nodata &  H em \\  
ZH 244(I1)(10584-19-1) & 9:55:12.78 68:59:45.74 & A I & 20.35  & 1,2,3 & var & low S/N, H$\alpha$ em\\ 
10584-9-1 & 9:55:18.97 69:08:27.54 & sgB[e] & 19.1 &  2 & var & H em P Cyg, He I, Fe II em, see text \\ 
ZH 364(I2)(10584-16-1) & 9:55:20.31 69:01:55.97 & LBVc\tablenotemark{b}  &  19.59 & 1,2,3 & var & H, He I, Fe II, NII em, H$\alpha$ wings, see text\\ 
ZH 235 &  9:55:22.52 68:58:32.85 & H II & 20.26  & 1 &  \nodata &  strong neb em, He I em \\ 
10584-5-2 & 9:55:25.61 69:12:14.00 &  F5 I & 19.98 & 2 & \nodata & neb em \\ 
ZH 224 (10584-23-1) & 9:55:34.51 68:55:48.50 & F 2-5 I & 18.83 & 1,2 & var: & strong neb em \\
10584-13-2 & 9:55:40.25 69:7:31.24   &  F2 I & 19.8  & 2  & \nodata &   H em \\ 
10584-10-5 & 9:55:41.24 69:11:02.53 & A I + WN & 19.65  & 2 & \nodata & blend, [N II] em + neb em \\ 
10584-20-2 & 9:55:53.35 68:59:04.49 & B5 I & 17.49  & 2 & \nodata & strong neb em, He I em, P cyg \\ 
10584-13-3 & 9:55:58.33 69:06:44.95 & F8 I & 19.62  & 2 & var: &  \\ 
ZH 354  &  9:56:01.36 68:59:49.37 & LBVc (Of/late WN:) & 19.6 & 1 & var: & H, strong N II, He II $\lambda$4686 em,  see text\\ 
10584-14-2 & 9:56:09.02 69:05:55.49 & G0 I & 19.65 & 2  & \nodata &  \\ 
ZH 1143 (10584-24-1) & 9:56:9.12 68:56:43.78 & F2 I & 18.76 & 1 & \nodata & H em, neb em, M81-75 \\  
ZH 1406(I8)(10584-18-1) & 9:56:14.76 69:05:19.88 & F5 I & 19.26  & 1,2,3 &  var: & neb em \\  
10584-25-2 & 9:56:15.70 68:58:32.63 & G0 I & 19.24  & 2 & var & neb em \\ 
ZH 348 & 9:56:24.51 68:59:15.91 & H II & 18.94  & 1 & \nodata & neb em, He I em \\
10584-18-5 & 9:56:32.36 69:05:08.99 & A8 I & 19.74  & 2 & \nodata &  \\ 

\enddata
\tablenotetext{a}{Primary Sources for targets: 1) \citet{ZH}, 2) GO-10584, 3) \citet{Sandage84a} } 
\tablenotetext{b}{Candidate LBV}
\end{deluxetable}

\begin{deluxetable}{lcccl}
\tabletypesize{\scriptsize}
\rotate
\tablenum{4}
\tablecolumns{5}
\tablecaption{Classification Comparison}
\tablewidth{0pc}
\label{tab:Comp}
\tablehead{\colhead{Star ID} & \colhead{This Paper} & \colhead{Previous Type} &  \colhead{Reference\tablenotemark{a}} & \colhead{Comments}}
\startdata  
NGC 2403 &     &     &     &    \\
ZH542  &  F V  &  H II  &  3   & neb em superposed \\ 
ZH553  &  A8 Ia & A5 Ia   &   1  &  IVa28, see text\\
ZH583  &  A:    & H II    & 3   & low S/N \\
ZH 585 &  F2 I  & LBVc\tablenotemark{b}    & 3   &          \\
ZH 730 &  A5    &  F0     & 1   & N2403-80 \citep{RMH87b}, High Vel, Horiz Br star:\\
ZH912  &  A0-2 I & A8 Ia  & 1   & neb em  \\
S29    &  A0     & LBVc   & 3   & broad H abs, neb em, prob foregrd \\
S44    & early F & LBVc   & 3   & low S/N \\
S94    &  F5 Ia &  LBVc   & 3   & see text  \\
S185   &   F8 V  & LBVc &   3   & narrow H abs, Horiz Br star:\\
V38    &  LBV   &  LBVc &  3   & see text \\
V52    &  F8 V   &  LBVc &  3   & see text \\
       &         &       &      &          \\
M81    &         &       &      &          \\
ZH224  &  F2-5 I   &  HII + blue cont. & 4 & neb em \\
       &  \nodata & LBVc   & 2  &       \\
ZH235  &  H II   &  H II  & 2  &        \\
ZH364(I2) &  LBVc &  LBV  & 2   & see text \\
ZH372  &  F5 I   &  SGc(F) &   2 &       \\
ZH479  &  F V    &  SNRc   &   2 &       \\
ZH491  &  foreground & SGc(G) & 2 & v.red, molecular bands? \\
ZH501  &  H II   &  H II  &  4 & strong neb em, He I em \\
       &  \nodata & H II  & 2  &       \\ 
ZH628  &  G V     &  G field & 4  &    \\
      &  \nodata &  FG field & 2  &     \\
ZH679 & A8 I     &  H II     & 2  & H em \\ 
ZH1143 & F2 Ia    &  F2 Ia   & 1  & M81-75, H em, neb em\\  
ZH1406(I8) & F5 I  &  LBVc   & 2  & see text \\
I3         &  F5 V &  LBVc   & 2  & see text \\
\enddata
\tablenotetext{a}{References for previous types: 1) \citet{RMH87b}, 2) \citet{Shol98a}, 3) \citet{Shol98b}, 4 \citet{Z96}  } 
\tablenotetext{b}{Candidate LBV}
\end{deluxetable}

\begin{deluxetable}{lllllllllllllll}
\tabletypesize{\scriptsize}
\rotate
\tablenum{5}
\tablecolumns{15}
\tablecaption{Multi-Wavelength Photometry\tablenotemark{a} }
\tablewidth{0pc}
\label{tab:Photshort}
\tablehead{
\colhead{Star ID} & 
\colhead{U\tablenotemark{b}} & 
\colhead{B\tablenotemark{b}} & 
\colhead{V\tablenotemark{b}}  & 
\colhead{R\tablenotemark{b}} &  
\colhead{F435\tablenotemark{c}} &
\colhead{F475\tablenotemark{c}} & 
\colhead{F606\tablenotemark{c}} & 
\colhead{1.1$\mu$m} & 
\colhead{1.6$\mu$m} & 
\colhead{3.6$\mu$m($\mu$Jy)} & 
\colhead{4.5$\mu$m($\mu$Jy)} &
\colhead{5.8$\mu$m($\mu$Jy)} &
\colhead{8$\mu$m($\mu$Jy)} & 
\colhead{Comments}
} 
\startdata 
      &     &     &    &    &    &  NGC 2403 &    &    &    &    &    &    &    &    \\
ZH 585\tablenotemark{d}  & 19.85 &  19.51 & 19.43  &  18.91 & \nodata  &  \nodata &  \nodata  &  \nodata  &  \nodata  &  13  &  1.6  &  \nodata  &  \nodata & \nodata \\
ZH 553                   & 18.24 &  18.21 & \nodata & 18.0  & \nodata  &  \nodata &  \nodata  &  \nodata  &  \nodata  &  38  &  12   &  \nodata  &  \nodata &  \nodata \\
ZH 335                   & 19.98 &   19.46&  19.19  &  19.05 & \nodata  &  \nodata &  \nodata  &  \nodata  &  \nodata  &  22 &  9    &  \nodata  &  \nodata & \nodata \\
ZH 2328\tablenotemark{d} & \nodata & \nodata & 20.04 & \nodata & \nodata & \nodata &  \nodata  &  \nodata  &  \nodata & \nodata & \nodata & \nodata & \nodata & \nodata\\
ZH 1521                  & 19.09 &  18.86 & 18.59 &   18.43  &  \nodata  &  \nodata & 19.98    &   \nodata  & \nodata  &  50  &  73  &  \nodata  &  \nodata &  \nodata\\
ZH 533                   & 18.66 &  19.24 & 19.36 &   19.39  &  \nodata  &  \nodata &  \nodata  & \nodata  &  \nodata  &  \nodata  &  \nodata  &  \nodata  &  \nodata &  \nodata\\
ZH 2313\tablenotemark{d} & \nodata & \nodata & 19.78 & \nodata & \nodata &  \nodata &  \nodata  &  \nodata  &  \nodata &  111  &  21  &  \nodata  &  \nodata &  complex region\\
ZH 2562\tablenotemark{d} & \nodata & \nodata & 19.57 & \nodata & \nodata &  \nodata &  \nodata  &  \nodata  &  \nodata  &  130  & 27  &  \nodata  &  \nodata &  H II? \\
ZH 2022\tablenotemark{d} & \nodata & \nodata & 19.95 & \nodata & \nodata &  \nodata &  \nodata  &  \nodata  &  \nodata  & \nodata  & \nodata  & \nodata  &  \nodata &  \nodata\\
10182-pr-9               & \nodata & \nodata & \nodata & \nodata &\nodata & 19.60   &  19.49    &  \nodata  &  \nodata  &  \nodata  & \nodata  & \nodata  & \nodata & \nodata\\ 
\enddata
\tablenotetext{a}{Only a portion of this table is shown here to demonstrate its form and content. A machine-readable version of the full table is available on-line.}
\tablenotetext{b}{Magnitudes from the LBT/LBC survey unless designated otherwise as a footnote to the star ID or in Comments. }
\tablenotetext{c}{Magnitude from HST images, see text.} 
\tablenotetext{d}{Photographic UBVR magnitudes from \citet{ZH}.} 
\end{deluxetable}

\begin{deluxetable}{lllllllll}
\tabletypesize{\scriptsize}
\rotate
\tablenum{6}
\tablecolumns{9}
\tablecaption{Interstellar Extinction and Luminosities (in magnitudes)}
\tablewidth{0pc}
\label{tab:Lum}
\tablehead{
\colhead{Star ID} & 
\colhead{Sp Type} & 
\colhead{$B-V$ (LBC)} & 
\colhead{$B-V$ (HST)}  & 
\colhead{E$_{BV}$} &  
\colhead{A$_{V}$} &
\colhead{M$_{V}$} & 
\colhead{M$_{Bol}$} &  
\colhead{Temp.\tablenotemark{a}}
} 
\startdata 
      &     &     & NGC 2403   &    &    &   &  &   \\
ZH 585  & F2 I &  0.08 & \nodata  &  \nodata & 0.5  &  -8.6 &  -8.5 & 7000   \\
ZH 553  & A8 I & \nodata & \nodata  &  \nodata & 0.5 & -9.6: & -9.5:& 8200   \\ 
ZH 335  & F0 I & 0.27 & \nodata &  0.07 &  0.22 &  -8.5 &   -8.4 & 7800     \\ 
ZH 1521  & F2 I & 0.27 & \nodata &  0.01: & 0.5 & \tablenotemark{b} & -8.0 & 7400 \\
ZH 533  & B5: I   & -0.12 & \nodata &  0.05: & 0.5 & -8.68 & -9.5 & 13700 \\
ZH 2313 &   A0-A2  &   \nodata  &  \nodata &  \nodata &  0.5 & -8.26 & -8.5  & 9500 \\
ZH 2562 & B I  &  \nodata  &  \nodata &  \nodata &  0.5 & -8.47  &  -10: & 20000:\\
ZH 2022 &  B8 I  &  \nodata  & \nodata  & \nodata  & 0.5  &  -8.09 & -8.6 & 10900 \\
10182-pr-9 &  WN  & \nodata & 0.11  & \nodata & \nodata &\nodata &  \nodata & \nodata \\ 
ZH 2016 & WN  & \nodata &  \nodata  & \nodata & \nodata & \nodata & \nodata & \nodata \\  
10182-pr-1 & A5-A8 I  &  \nodata  & 0.36  & 0.24  & 0.77  & -9.17  & -9.1 & 8300  \\
ZH 946    &  B5 I:  &  0.05  &  \nodata & 0.0:  & 0.5    & -8.3   &  -9.3 & 13700 \\    
ZH 947    & B5 I  &  -0.21 &  \nodata & \nodata & 0.5  & -8.98  &  -10.0 & 13700 \\ 
10182-pr-2\tablenotemark{c} & F5 I   &  1.33  & 1.28 &  1.00 &  3.2  & -11.7:  & -11.6 &  7000 \\ 
10182-pr-6 & A8-F0 I & 0.89  & 1.00 & 0.7-0.8 & 2.4 & -10.4:  & -10.3 & 8000  \\
10182-pr-15 &  A0-A2 I & 0.17 & 0.35 & 0.12-0.30 & 0.96 & -8.6 & -8.8 & 9500  \\
10182-pr-16 & B8 I: & -0.14 & 0.08 & 0.05 & 0.16 & -7.7 &  -8.2 & 10900 \\
ZH 729  &  B8 I &  0.15 &  \nodata & 0.20  & 0.64 & -8.9  &  -9.4 & 10900 \\
ZH 2248 &  A0 I  & \nodata  & \nodata & \nodata & 0.5 & -8.3 & -8.6 & 9800  \\
V38  &  LBV & -0.02 &  \nodata  & \nodata  & 0.5  & -8.6     & -10.25 & (18950)  \\
ZH 931  & A5-A8 I  & 0.27 &  \nodata  & 0.15   & 0.48 & -8.0  & -8.1 &  8300 \\
ZH 1938 & A2 I & \nodata &  \nodata & \nodata & 0.5  & -8.7 &  -8.9 &  9100 \\ 
S94 & F5 I  & 0.53 & \nodata & 0.20  & 0.64  & -8.8   &  -8.8 & 7000 \\ 
ZH 924 &  B5-B8 I  & -0.13  &  \nodata  & 0.0:  & 0.5  & -8.7 & -9.4 &  12000 \\ 
ZH 912 &  A0-A2 I  & 0.2  & \nodata  &  0.15 & 0.48   &  -9.6 &  -9.8 & 9500 \\ 
ZH 884  & F0 I  & 0.07  &  \nodata  &  \nodata & 0.5 &  -9.3  &  -9.3 & 7800 \\ 
      &     &     &    &    &    &   &  &  \\ 
      &     &     &  M81  &    &    &   &  &   \\  
10584-11-1 & A5 I  &  0.27 & 0.35 & 0.25  & 0.8   & -9.1   & -9.1 & 8500  \\  
ZH 679   & A8 I   &  0.27 &  \nodata  & 0.13 & 0.42 & -8.5 & -8.4 & 8200 \\
10584-8-4 & sgB[e]  & \nodata & 0.25 & \nodata & (0.9) &  \nodata  & -9.4 & (21700) \\
10584-4-1 & sgB[e]  &  0.04   & 0.36  &  \nodata & (0.9) & \nodata &  -10.7 & (19671)  \\
ZH 372(10584-15-1)  & F5 I  & 0.38 &  0.50 &  0.17 &  0.54 &  -8.8 & -8.7 & 7000 \\
10584-8-1  & F0 I & 0.45  & 0.53 & 0.25--0.33 & 0.8--1.06 & -9.5  & -9.4 & 7800 \\ 
10584-8-2  & F0 I  & 0.41 & 0.49 & 0.21--0.29 & 0.7--0.9  & -9.1 & -9.0 &  7800 \\  
ZH 244(I1)(10584-19-1) & A I & 0.04 & 1.29: & 0 -? &  (0.9)  & -9.1 & -9.1 & 9000: \\  
10584-9-1  &  sgB[e] & 0.24 & 0.35  & \nodata & (0.9)  & \nodata & -11.0 & (18000)  \\ 
ZH 364(I2)(10584-16-1) & LBVc & 0.27 & 1.54 & \nodata & (0.9)  &  \nodata & -10.4 &  (15860)  \\  
10584-5-2  & F5 I  & 0.23 & 0.38 & 0.05 & 0.16 & -7.9 & -7.8 & 6200  \\
ZH 224 (10584-23-1) & F2-5 I & 0.47 & 1.58: & 0.22 & 0.70 & -9.7 & -9.6 & 7200 \\ 
10584-13-2 &  F2 I  & 0.69 & 0.69 & 0.43 & 1.38 & -9.4 & -9.3 & 7400  \\
10584-10-5 & A I + WN  & 0.10 & 0.38 & \nodata &  \nodata  & \nodata  & \nodata & \nodata  \\
10584-20-2 & B5 I  & 0.86:  & 0.20 & 0.28 &  0.90 &  -9.2:  & -10.0 & 13700 \\
10584-13-3  & F8 I  & 1.36 & 1.27 & 0.8 & 2.56 & -10.3: & -10.2 &  6200 \\
ZH 354  &  LBVc  &  0.1 & \nodata & \nodata & (0.9) &  \nodata & \nodata & \nodata  \\
10584-14-2 & G0 I  & 0.78 & 0.81 & 0.20 & 0.64 & -8.7 &  -8.7 &  5500 \\
ZH 1143 (10584-24-1)  & F2 I  & 0.42   & 1.42:   &  0.16  &  0.51 & -9.55 & -9.45 & 7400 \\
ZH 1406(I8)(10584-18-1) & F5 I & 0.51 & 0.61 & 0.23 &  0.74  & -9.2 & -9.1 & 7000 \\
10584-25-2 & G0 I  & 1.0   & 1.0 &  0.35 &  1.12 &  -9.7  &  -9.7 &  5500 \\
10584-18-5 & A8 I  & 0.16  &  0.27  & 0.13 &  0.41 & -8.3 & -8.2 &  8200 \\
\enddata
\tablenotetext{a}{Temperatures based on the Planck fit to the SED are in parenthses} 
\tablenotetext{b}{M$_{v}$ depends on the adopted visual magnitude: -9.45(LBC),-8.06(HST).} 
\tablenotetext{c}{Note exceptionally high luminosity. Its HST image in Figure B1 shows it is extended.}
\end{deluxetable}

\begin{deluxetable}{llccccl}
\tabletypesize{\scriptsize}
\rotate
\tablenum{A1}
\tablecolumns{7}
\tablecaption{Foreground Stars and Others}
\tablewidth{0pc}
\label{tab:Others}
\tablehead{\colhead{Star ID} & \colhead{Position${J2000}$}
  & \colhead{Sp. Type} & \colhead{$V$} &  \colhead{Source} &  \colhead{Variability} & \colhead{Comments}}
\startdata  
NGC 2403     &                       &     &        &         &   &                    \\
ZH 2141 (F2) & 7:35:13.38 65:37:2.01 &  A2 & 201.13  & 1 & \nodata & Horiz. br star:\\ 
ZH 601 (F1+F2) & 7:35:22.74 65:37:26.30 & F5 V & 19.66  & 1 & \nodata &  \nodata\\
ZH 1882 (F1+F2) & 7:35:22.83 65:35:21.09 & \nodata & 19.97  & 1 & \nodata &  low S/N \\
ZH 608 (F2) & 7:35:22.94 65:39:9.78 & F5 V & 18.41 &  1 & \nodata & \\ 
ZH 604 (F1) & 7:35:29.94 65:39:20.71 & pec &  19.56 & 1 & var: & pec em, low S/N \\
ZH 803 (F2) & 7:35:36.57 65:39:20.57 & A5 & 201.19 &  1 & \nodata & low S/N\\ 
ZH 360 (F1+F2) & 7:35:37.63 65:33:50.97 & \nodata & 20.05  & 1 & \nodata & low S/N\\
ZH 583 (F2) & 7:35:41.00 65:35:59.15 & A: & 19.91  & 1 & \nodata & low S/N\\
ZH 2125 (F1) & 7:35:42.92 65:37:44.33 & \nodata &  19.5  & 1 &  \nodata & low S/N, foreground\\
ZH 593 (F2) & 7:35:43.26 65:38:21.20 & A: & 20.12  & 1 & \nodata & low S/N, foreground\\
ZH 2131 (F1) & 7:35:43.51 65:38:39.81 & \nodata &  19.31  & 1 & \nodata & low S/N\\ 
ZH 790 (F1+F2) & 7:35:53.66 65:39:35.24 & A8 & 19.77 & 1 & \nodata  & \nodata \\ 
ZH 569 (F1) & 7:35:53.76 65:34:53.01 &  \nodata & 20.08 & 1 & \nodata & neb em superposed \\ 
ZH 190 (F1+F2) & 7:35:58.27 65:29:43.74 & A2 & 19.35   & 1 & \nodata & Horiz. br. star:\\ 
ZH 1005 (F1+F2) & 7:35:58.86 65:44:55.60 & A5 & 18.31  & 1 & var: & Horiz. br. star:\\
ZH 1001 (F1+F2) & 7:35:59.71 65:43:01.14 & F5 V & 17.97  & 1 & \nodata & High vel star \\
ZH 566 (F2)    & 7:36:00.91 65:35:16.73 & \nodata & 19.86  & 1 &  \nodata & v low S/N, pec\\ 
ZH 565 (F1+F2) & 7:36:04.24 65:35:48.09 &  \nodata & 19.56  & 1 & var: & pec:, low S/N\\
ZH 991 (F2) & 7:36:07.44 65:41:55.23 & F5 V & 19.58  & 1 & \nodata &  \\ 
ZH 1819 (F1+F2) & 7:36:20.17 65:28:42.21 & F5 V & 17.97  & 1 & \nodata &  \nodata\\
ZH 552 (F2) & 7:36:24.26 65:35:51.06 & A5 & 19.64  & 1 & \nodata & neb em, H em superposed \\ 
ZH 542 (F1+F2) & 7:36:27.42 65:33:49.64 & F V & 20.07  & 1 & \nodata & neb em superposed \\ 
ZH 1100 (F1+F2) & 7:36:32.43 65:43:56.36 & \nodata & 20.2  & 1 & \nodata & low S/N, foreground\\ 
S44 (F2) & 7:36:38.96 65:35:32.27 &  early F & 19.59  & 5 & \nodata & low S/N \\ 
ZH 1097 (F1+F2) & 7:36:42.74  65:44:9.38 & F8 V & 18.73 & 1 & \nodata & \nodata\\
ZH 732 (F2) & 7:36:45.02 65:38:50.60 & F5 V & 20.05  & 1 & \nodata & neb em \\ 
10182-p2-22 (F2) & 7:36:49.63 65:36:22.57 & G: V & 20.08  & 2 & \nodata & H em superposed\\ 
10579-x1-2 (F2) & 7:36:49.79 65:35:49.69 & G: V: & 19.58  & 3 & \nodata &  neb em, H em superposed\\ 
V52 (F1) & 7:36:50.39 65:37:52.1 & F8 V &  20.10  & 4 &  \nodata & \\ 
S29 (F2) & 7:36:52.35 65:34:53.58 & early A & 18.89 & 5 & \nodata & br H abs, neb em superposed \\ 
ZH 2276 (F2) & 7:36:56.99 65:38:1.57 & F0 V & 19.68  & 1 & \nodata & neb em \\
ZH 730 (F1) & 7:37:00.29 65:37:55.25 & A5 & 18.16  & 1 & \nodata & High vel star, Hor.Br?, N2403-80 \citep{RMH87b} \\ 
S185 (F2) & 7:37:02.42 65:35:54.64 & F8 V & 17.84  & 5 & \nodata & narrow H abs., Horiz Br? \\ 
ZH 1081 (F1) & 7:37:08.38 65:42:13.64 &  \nodata & 18.98 & 1 & \nodata & low S/N, pec \\
ZH 292 (F1+F2) & 7:37:11.31 65:28:32.69 & G0 V & 18.74 & 1 & \nodata  & \nodata\\ 
ZH 932 (F2) & 7:37:13.49  65:40:18.07 & A: & 19.91  & 1 & \nodata & low S/N \\  
ZH 923 (F2) & 7:37:21.05  65:39:11.48 & A8 & 18.1  & 1 & \nodata & Horiz Br? \\ 
ZH 1079 (F1+F2) & 7:37:21.35 65:42:47.77 &  \nodata & 18.68  & 1 & \nodata & low S/N \\ 
ZH 480 (F2) & 7:37:21.62 65:32:32.53 & A5 & 19.75  & 1 & \nodata &  \nodata \\ 
ZH 2600 (F1+F2) & 7:37:24.04 65:40:46.79 & F2-F5 V & 19.49  & 1 & \nodata & \nodata\\ 
ZH 1064 (F1+F2) & 7:37:48.52 65:40:53.14 & F5-F8 V & 18.8   & 1 & \nodata & \nodata\\
ZH 897 (F1) & 7:37:50.46 65:38:35.70 & late A & 19.32  & 1 & \nodata & low S/N \\ 
ZH 1061 (F2) & 7:37:56.58 65:39:15.43 & foreground & 18.94  & 1 & \nodata & red only \\ 
ZH 2455 (F1+F2) & 7:37:59.50 65:37:25.69 & F2 V & 19.58  & 1 & \nodata & \nodata\\ 
ZH 418 (F1+F2) & 7:38:9.62 65:26:44.31 & \nodata & 19.51 & 1 & \nodata & low S/N\\
ZH 869 (F1+F2) & 7:38:23.77 65:36:38.69 & QSO &  19.74 & 1 & var &  \nodata \\ 
               &                        &     &        &      &   &              \\
M81            &                        &     &        &      &   &              \\
ZH 400         & 9:54:18.65  69:02:45.65 & galaxy & 19.43 &  1 & \nodata & red shifted \\
ZH 512         & 9:54:20.69  69:09:13.09 & \nodata & 19.62  & 1  & \nodata & poor S/N \\ 
10584-3.1      & 9:54:27.43 69:08:59.07 & \nodata & 19.97  & 6 & \nodata & low S/N \\ 
ZH 1355        & 9:54:34.94 69:01:54.92 & F: V & 19.75  & 1 & \nodata & neb em, low S/N \\ 
ZH 833         & 9:54:50.64 69:14:44.83 & F5 V  & 19.85  & 1 & \nodata & Horiz Br? \\ 
ZH 491         & 9:54:56.81 69:4:55.16 & \nodata & 19.63 & 1 & \nodata& v.red, molecular bands\\ 
ZH 1389        & 9:55:00.11  69:06:14.05 & F V & 19.88  & 1 & \nodata & neb em \\ 
ZH 904         & 9:55:03.60  69:16:43.73 & A5  & 19.72  & 1 & \nodata & Horiz Br \\ 
ZH 1209        & 9:55:15.78 69:5:47.14 & F2 V  & 19.14 & 1 & \nodata & \nodata \\ 
ZH 228         & 9:55:42.39 68:58:22.39 & F: V & 19.99 & 1 & \nodata & low S/N \\ 
ZH 108         & 9:55:44.55 68:51:52.70 & neb em & 20.09  & 1 & \nodata & red sp. only \\ 
ZH 619         & 9:55:51.71 69:04:40.96 & F8  V & 20.03  & 1 & \nodata & \nodata \\ 
ZH 628         & 9:55:51.90 69:07:39.05 & G V & 18.66  & 1 & \nodata & \nodata \\ 
10584-10.4     & 9:55:51.98 69:12:9.57 & F8 V & 19.41  & 6 & \nodata &  Horiz Br star \\
10584-21.4     & 9:55:55.03 69:00:56.28 & F5 V & 19.61  & 6 &  \nodata   &  \nodata \\
10584-10.1     & 9:56:2.63 69:11:45.27  & G V & 19.44   & 6 & \nodata &  \nodata \\
ZH 623         & 9:56:03.22 69:08:09.43 & F5 V & 18.48  & 1 & \nodata & Horiz Br? \\ 
I3             & 9:56:08.50 69:03:51.24 & F5 V & 19.6  & 7 & \nodata & LBV cand., see text \\ 
10584-10.2     & 9:56:11.26 69:10:47.26 & F V & 19.51  &  6 & \nodata &  \nodata \\
ZH 479         & 9:56:20.54 69:2:48.46 & F V & 19.14  &  1 & \nodata &  \nodata \\  
10584-21.8     & 9:56:24.16 69:00:29.28 & F V & 19.91  & 6 &  \nodata &  \nodata \\
10584-21.2     & 9:56:27.52 69:01:10.04  & F5 V & 19.52 & 6 &  \nodata & Horiz Br  \\ 
10584-21.5     & 9:56:27.55 69:01:9.95   & F V  & 19.65  & 6 &  \nodata &  red sp. only \\
ZH 642         & 9:56:33.06 69:08:02.69  & \nodata & 19.1  & 1 & \nodata & poor S/N\\ 
ZH 344         & 9:56:33.07 68:58:30.15  & A  & 19.24  & 1  & \nodata & Horiz Br  \\ 
ZH 92          & 9:56:35.17 68:50:45.28 & A-F V & 19.94  & 1 & \nodata &  \nodata \\
10584-22.3     & 9:56:36.47 69:00:28.78  & QSO & 20.69  & 6 &  \nodata &  \nodata \\ 
10584-22.1     &  9:56:49.54 69:03:13.11 &  A-type WD & 19.8  & 6 & \nodata &  \nodata \\
ZH 865        & 9:56:53.05 69:10:27.22 & G V & 19.59  & 1 & \nodata  &  \nodata \\ 
ZH 454        & 9:56:58.29 69:00:45.92 & pec em QSO? & 19.93 & 1 & \nodata &  \nodata \\ 
ZH 324        & 9:57:01.56 68:55:0.29 & QSO: & 17.87 & 1 &  \nodata & pec em \\ 
\enddata
\tablenotetext{a}{Primary Sources for targets: 1) \citet{ZH}, 2) GO-10182(PI), 3) GO-10579(PI), 4) \citet{ST68}, 5) \citet{Sandage84a}, 6) GO-10584(PI), 7) \citet{Sandage84b} } 
\end{deluxetable}

\begin{deluxetable}{lllllllllllllll}
\tabletypesize{\scriptsize}
\rotate
\tablenum{5}
\tablecolumns{15}
\tablecaption{Multi-Wavelength Photometry}
\tablewidth{0pc}
\label{tab:Phot}
\tablehead{
\colhead{Star ID} & 
\colhead{U\tablenotemark{a}} & 
\colhead{B\tablenotemark{a}} & 
\colhead{V\tablenotemark{a}}  & 
\colhead{R\tablenotemark{a}} &  
\colhead{F435\tablenotemark{b}} &
\colhead{F475\tablenotemark{b}} & 
\colhead{F606\tablenotemark{b}} & 
\colhead{1.1$\mu$m} & 
\colhead{1.6$\mu$m} & 
\colhead{3.6$\mu$m($\mu$Jy)} & 
\colhead{4.5$\mu$m($\mu$Jy)} &
\colhead{5.8$\mu$m($\mu$Jy)} &
\colhead{8$\mu$m($\mu$Jy)} & 
\colhead{Comments}
} 
\startdata 
      &     &     &    &    &    &  NGC 2403 &    &    &    &    &    &    &    &    \\
ZH 585\tablenotemark{c}  & 19.85 &  19.51 & 19.43  &  18.91 & \nodata  &  \nodata &  \nodata  &  \nodata  &  \nodata  &  13  &  1.6  &  \nodata  &  \nodata & \nodata \\
ZH 553                   & 18.24 &  18.21 & \nodata & 18.0  & \nodata  &  \nodata &  \nodata  &  \nodata  &  \nodata  &  38  &  12   &  \nodata  &  \nodata &  \nodata \\
ZH 335                   & 19.98 &   19.46&  19.19  &  19.05 & \nodata  &  \nodata &  \nodata  &  \nodata  &  \nodata  &  22 &  9    &  \nodata  &  \nodata & \nodata \\
ZH 2328\tablenotemark{c} & \nodata & \nodata & 20.04 & \nodata & \nodata & \nodata &  \nodata  &  \nodata  &  \nodata & \nodata & \nodata & \nodata & \nodata & \nodata\\
ZH 1521                  & 19.09 &  18.86 & 18.59 &   18.43  &  \nodata  &  \nodata & 19.98    &   \nodata  & \nodata  &  50  &  73  &  \nodata  &  \nodata &  \nodata\\
ZH 533                   & 18.66 &  19.24 & 19.36 &   19.39  &  \nodata  &  \nodata &  \nodata  & \nodata  &  \nodata  &  \nodata  &  \nodata  &  \nodata  &  \nodata &  \nodata\\
ZH 2313\tablenotemark{c} & \nodata & \nodata & 19.78 & \nodata & \nodata &  \nodata &  \nodata  &  \nodata  &  \nodata &  111  &  21  &  \nodata  &  \nodata &  complex region\\
ZH 2562\tablenotemark{c} & \nodata & \nodata & 19.57 & \nodata & \nodata &  \nodata &  \nodata  &  \nodata  &  \nodata  &  130  & 27  &  \nodata  &  \nodata &  H II? \\
ZH 2022\tablenotemark{c} & \nodata & \nodata & 19.95 & \nodata & \nodata &  \nodata &  \nodata  &  \nodata  &  \nodata  & \nodata  & \nodata  & \nodata  &  \nodata &  \nodata\\
10182-pr-9               & \nodata & \nodata & \nodata & \nodata &\nodata & 19.60   &  19.49    &  \nodata  &  \nodata  &  \nodata  & \nodata  & \nodata  & \nodata & \nodata\\ 
ZH 2016\tablenotemark{c} & \nodata & \nodata & 18.47 &  \nodata  & \nodata & \nodata & \nodata  & \nodata  & \nodata  & \nodata  &  \nodata  &  \nodata  &  \nodata &  \nodata\\ 
10579-x1-3               &  18.72 & 19.15 & 20.3 &  19.39 & \nodata &   19.78 & 19.58  &  17.72 &  16.85  &  \nodata  &  \nodata  &  \nodata  &  \nodata &  \nodata\\ 
10182-pr-1               &  17.65 & 18.01 & \nodata & \nodata &\nodata & 19.23 & 18.87 & \nodata & \nodata &  \nodata &  \nodata  &  \nodata  &  \nodata & \nodata \\
ZH 946                   &  19.64 & 19.79 & 19.74   & 19.57   & \nodata & \nodata & \nodata & 19.27 & 18.55 & \nodata  &  \nodata  &  \nodata  &  \nodata &  \nodata\\    
ZH 947                   &  18.17 & 18.85 & 19.06   & 19.1    & \nodata & \nodata & \nodata & \nodata & \nodata & \nodata & \nodata  &  \nodata  &  \nodata & \nodata  \\ 
10182-pr-2               &  20.5  & 20.16 & 18.83   & 18.3    & \nodata & 20.30 & 19.02     & 17.25    & 16.78            & 65  &  46   &  15   &  \nodata & \nodata \\ 
10182-pr-6               & 20.09 &  19.68  &  18.79 & 18.52   &  \nodata & 20.52 & 19.52    & 18.04     & 17.65           & 614   & 376   &  fuzzy  & \nodata & blended, H II?   \\
10182-pr-15              &  19.34 & 19.69 & 19.52 & 19.39     &  \nodata &  20.22 & 19.87   &  19.2     & 19.2            & \nodata & \nodata & \nodata & \nodata &  \nodata \\
10182-pr-16              &  18.85 &  19.21 & 19.35 & 19.35    & \nodata &  20.0  & 19.92    & \nodata   & \nodata          & \nodata & \nodata & \nodata & \nodata &  \nodata \\
ZH-729                   &  19.23 & 19.36 & 19.21 & 19.02     & \nodata & \nodata & \nodata & 19.49      & 18.45         & 65   & \nodata & \nodata & \nodata &  \nodata \\
V37                      &  20.85 & 21.62 & 19.97 & 20.01     & \nodata & \nodata & \nodata & \nodata    & \nodata      &\nodata & \nodata & \nodata & \nodata & LBV, see \citet{RMH2017} \\
ZH 2248\tablenotemark{c} & \nodata & \nodata & 19.71 & \nodata & \nodata & \nodata & \nodata & \nodata   & \nodata      & \nodata & \nodata & \nodata & \nodata & \nodata \\
V38                      &  18.49 & 19.39 & 19.41 & 19.21    & \nodata   & \nodata  & \nodata & 21.1     & \nodata      & \nodata & \nodata & \nodata & \nodata & LBV see text \\
ZH 931\tablenotemark{c}  & \nodata & 20.22 & 19.95 & 19.66   & \nodata  & \nodata   & \nodata & \nodata  & \nodata      & \nodata & \nodata & \nodata & \nodata & \nodata  \\
ZH 1938\tablenotemark{c} & \nodata & \nodata & 19.36 & \nodata & \nodata & \nodata  & \nodata & \nodata  & \nodata      & \nodata & \nodata & \nodata & \nodata & \nodata \\ 
S94                      & \nodata & \nodata & \nodata &\nodata &\nodata & \nodata  & \nodata & 17.77    & 17.57        & 45     & 16  &  \nodata & \nodata & visual photometry\tablenotemark{d} \\ 
ZH 924                   & 18.61  & 19.17 & 19.3 & 19.32     & \nodata  & \nodata  & \nodata  & \nodata & \nodata       & 3       &  \nodata & \nodata & \nodata & \nodata  \\ 
ZH 912                   & 18.3 & 18.61 & 18.41 & 18.25      & \nodata  &  \nodata & \nodata  &  \nodata & \nodata       & 182  & 46  &  \nodata & \nodata & \nodata  \\ 
ZH 884                   & 19.12 & 18.85 & 18.78 & 18.67    & \nodata  &  \nodata & \nodata &  \nodata & \nodata & \nodata & \nodata & \nodata & \nodata & \nodata  \\ 
      &     &     &    &    &    &   &    &    &    &    &    &    &    &    \\
      &     &     &    &    &    &  M81  &    &    &    &    &    &    &    &   \\  
10584-11-1               & 19.32  &  19.65 & 19.38 & 19.32  & 19.88   & \nodata   & 19.53   &  18.49  & \nodata    & \nodata    & \nodata     & \nodata    & \nodata   \\  
ZH 679                   & 20.11  &  19.97 & 19.7  & 19.56   & \nodata   & \nodata & \nodata & \nodata & \nodata & \nodata    & \nodata    & \nodata     & \nodata    & \nodata   \\
10584-8-4                & 18.87:  & \nodata & \nodata & \nodata & 20.14 &  \nodata  &  19.89 & \nodata & \nodata &  392    & 411     &  212           &   129        &  \nodata   \\
10584-4-1                & 19.07  &  19.72 &  19.68 & 19.55 &  19.86 &  \nodata &   19.50 & \nodata & \nodata  & \nodata    & \nodata    & \nodata     & \nodata    & \nodata   \\
ZH 372(10584-15-1)       & 20.36 & 20.02 &  19.64 &  19.39 &  20.06 &  \nodata &   19.56  &  \nodata & 18.43 &  \nodata    & \nodata    & \nodata     & \nodata    & \nodata   \\
10584-8-1                &  19.71 & 19.61 & 19.16 & \nodata & 19.79 & \nodata &  19.26  &   \nodata  & \nodata  &   101  &   165      &  76     &  282:          &  \nodata   \\ 
10584-8-2                &  20.16 & 19.85 & 19.46 & 19.2    &  20.00 & \nodata &  19.51  &   \nodata  & \nodata  &  70  &  36         &  fuzzy   & \nodata    & \nodata   \\  
ZH 244(I1)(10584-19-1)   &  19.54 & 20.39 & 20.35 & 20.23 &  20.86 & \nodata &  19.57 & \nodata & \nodata  & \nodata    & \nodata    & \nodata     & \nodata    & \nodata   \\  
10584-9-1                &  18.45 &  19.34 & 19.1  & 18.87 & 19.49 &  \nodata & 19.14  & \nodata &  \nodata &  175    &  102         &   1205:: &  \nodata    & HII contam.   \\ 
ZH 364(I2)(10584-16-1)   & 18.84 & 19.86 &  19.59 & 19.09 &  20.90 &  \nodata & 19.36  &  \nodata & 18.23 &  38    & 36     &  fuzzy &  fuzzy &  \nodata   \\  
10584-5-2                & 20.66 & 20.21 & 19.98 & 19.84 & 20.46 & \nodata &  20.08  & \nodata & 19.24 & \nodata & \nodata & \nodata  & \nodata    & \nodata \\
ZH 224 (10584-23-1)      & 19.54 & 19.31 & 18.83 & 18.63 & 20.90 & \nodata & 19.42 & \nodata  &  \nodata  & 46   & 6   & \nodata &  \nodata & \nodata \\ 
10584-13-2               &  21.09 &  20.49 & 19.8 & 19.5 &  20.45 & \nodata &  19.76  & \nodata   &  18.6 &  \nodata  &  \nodata  & \nodata  &  \nodata &  \nodata \\
10584-10-5               & 19.31 &   19.75 &  19.65 & 19.47  & 20.11 & \nodata & 19.73  &  \nodata  & \nodata  &  \nodata  & \nodata  &  \nodata &  \nodata & \nodata \\
10584-20-2               & 16.6  & 18.35 & 17.49 &  18.34 &  19.71   & \nodata &  19.51  &  \nodata & \nodata & \nodata  & \nodata  &  \nodata  & \nodata & \nodata  \\
10584-13-3               & 22.15 & 20.98 & 19.62 & \nodata & 21.19 & \nodata &  19.92  &  \nodata &  \nodata & \nodata & \nodata  & \nodata  &  \nodata  & \nodata \\
ZH 354                   &  19.3 & 19.7 & 19.6 & 19.46 & \nodata & \nodata & \nodata &  \nodata & 18.7 & \nodata &  \nodata &  \nodata  &  \nodata  & \nodata \\
10584-14-2               & 21.14 & 20.43 & 19.65 & 19.38 & 20.56 & \nodata &  19.75  &  \nodata &  \nodata & \nodata &  \nodata &  \nodata  &  \nodata  & \nodata \\
ZH 1143 (10584-24-1)     & 19.48   & 19.18   & 18.76   &  18.65  &  20.54  & \nodata &  19.12  &  \nodata &  \nodata & \nodata &  \nodata &  \nodata  &  \nodata  & \nodata \\
ZH 1406(I8)(10584-18-1   & 20.27 & 19.75 &  19.26 & 19.07 &  19.98   &   \nodata & 19.37  &  \nodata & 18.16 & \nodata &  \nodata &  \nodata  &  \nodata  & \nodata \\
10584-25-2               & 20.42  & 20.25   & 19.24  &  18.93  &  20.26 &  \nodata &  19.26  &  \nodata & 17.6 &  43   &  14  &   \nodata  &  \nodata  & \nodata \\
10584-18-5               & 20.08  & 19.9  &  19.74  & 19.7 &  20.17 & \nodata &  19.90   &  \nodata &  19.24 &   \nodata &  \nodata &  \nodata  &  \nodata  & \nodata \\
\enddata
\tablenotetext{a}{ Magnitudes from LBT/LBC unless designated otherwise as a footnote to the star ID or in Comments. } 
\tablenotetext{b}{Magnitudes from HST images, see text.} 
\tablenotetext{c}{Photographic UBVR magnitudes from \citet{ZH}.} 
\tablenotetext{d}{V=19.31, B-V=0.53 \citet{Sandage84b}}
\end{deluxetable}



\begin{thebibliography}{}

\bibitem[Aret et al.(2016)]{Aret}Aret, A., Kraus, M., \& Slecha, M. 2016, \mnras, 456, 1424

\bibitem[Cardelli et al.(1989)]{Cardelli}Cardelli, J. A., Clayton, G. C., \& 
  Mathis, J. S. 1989, \apj, 345, 245 

\bibitem[Ekstrom et al.(2012)]{Ekstrom}Ekstrom, S. et al. 2012, \aap, 537, A146 

\bibitem[Fabricant et al(1998)]{Fab98}Fabricant, D. G., Hertz, E. N., Szentgyorgyi, A. H., et al.  1998, Proc. SPIE, 3355, 285

\bibitem[Fabricant et al.(2005)]{Fab}Fabricant, D., Fata, R., Roll, J., et al. 2005, \pasp, 117, 1411

\bibitem[Flower(1996)]{Flower}Flower, P. J. 1996, \apj, 469, 355


\bibitem[Freedman et al.(2001)]{Freedman2001}Freedman, W. L., Madore, B. F., Gibson, B. K. et al. 2001, \apj, 553, 47 

\bibitem[Gerke, Kochanek \& Stanek(2014)]{Gerke}Gerke, J. R., Kochanek, C. S. \&
 Stanek, K. Z. 2014, \mnras, 450, 3289  

\bibitem[Gordon, Humphreys \& Jones(2016)]{Gordon}Gordon, M. S., Humphreys, R. M, \& Jones, T. J. 2016, 825, 50 

\bibitem[Grammer, Humphreys \& Gerke(2015)]{Grammer}Grammer, S. H., Humphreys, R. M, \& Gerke, J. 2015, \aj, 149, 152  

\bibitem[Humphreys \& Davidson(1979)]{HD79}Humphreys, R. M. \& Davidson, K. 1979, \apj, 232, 409

\bibitem[Humphreys(1980)]{RMH80}Humphreys, R. M. 1980, \apj, 241, 598

\bibitem[Humphreys et al.(1986)]{RMH86}Humphreys, R. M., Aaronson, M., Lebofsky, M. et al. 1986, \aj, 91, 808 
\bibitem[Humphreys \& Aaronson(1987a)]{RMH87a}Humphreys, R. M. \&  Aaronson, M. 1987a, \apjl, 381L, 69 

\bibitem[Humphreys \& Aaronson(1987b)]{RMH87b}Humphreys, R. M. \&  Aaronson, M. 1987b, \aj, 94, 1156 

\bibitem[Humphreys \& Davidson(1994)]{HD94}Humphreys, R. M. \& Davidson, K. 1994, \pasp, 106, 1025
\bibitem[Humphreys et al.(2013)]{RMH13}Humphreys, R. M., Davidson, K., \& Grammer, S. H. 2013, \apj, 773, 46  

\bibitem[Humphreys et al.(2016)]{RMH16}Humphreys, R. M., Weis, K., Davidson, K. \& Gordon, M. S. 2016, \apj, 825, 64   

\bibitem[Humphreys et al(2017a)]{RMH.4}Humphreys, R. M., Gordon, M. S., Martin, J. C., Weis, K. \& Hahn, D. 2017a, \apj, 836, 64  

\bibitem[Humphreys et al(2017b)]{RMH.5}Humphreys, R. M., Davidson, K., Hahn, D., Martin, J. C., Weis, K. 2017b, \apj, 844, 40  

\bibitem[Humphreys et al(2017c)]{RMH2017}Humnphreys, R. M., Davidson, K. Van Dyk, S. D., \& Gordon, M. S. 2017c, \apj, 848, 86   

\bibitem[Kochanek et al.(2008)]{CSK}Kochanek, C. S., Beacom, J. F., Kistler, M. D. et al. 2008, \apj, 684, 1336  


\bibitem[Kraus et al.(2014)]{Kraus}Kraus, M., Cidale, L. S., Arias, M. L., Oksala, M. E. \& Borges Fernandes, M. 2014, \apjl, 780, L10

\bibitem[Kudritzki et al.(2012)]{Kud}Kudritzki, R-P., Urbaneja, M. A., Gazak, Z., et al. 2012, \apj, 747, 15  

\bibitem[Martins et al.(2005)]{Martins}Martins, F., Schaerer, D. \& Hillier, D. 
J. 2005, \aap, 436, 1049 


\bibitem[Sandage(1984a)]{Sandage84a}Sandage, A. 1984a, \apj, 89, 621 

\bibitem[Sandage(1984b)]{Sandage84b}Sandage, A. 1984b, \apj, 89, 630 

\bibitem[Sholukhova et al.(1998a)]{Shol98a}Sholukhova, O. N., Fabrika, S. N., Vlasyuk, V. V., \& Dodonuv, S. N. 1998a, Pis'ma v Astronomicheskii Zhurnal, 24, 591

\bibitem[Sholukhova et al.(1998b)]{Shol98b}Sholukhova, O. N., Fabrika, S. N., \& Vlasyuk, V. V., 1998b,  Pis'ma v Astronomicheskii Zhurnal, 24, 699  

\bibitem[Smith \& Tombleson(2015)]{Smith2015}Smith, N. \& Tombleson, R. 2015, \mnras,447, 598 

\bibitem[Tammann \& Sandage(1968)]{ST68}Tammann, G. A. \& Sandage, A. 1968, \apj, 151, 825 

\bibitem[Van Dyk et al.(2006)]{vandyk}Van Dyk, S. D, Li, Weidong, Filippenko,
 A. V., Humphreys, R. M., Chornock, R.,, Foley, R., \& Challis, P. M. 2006,
 arXiv:astro-ph/0603025v1

\bibitem[Zickgraf \& Humphreys(1991)]{ZH}Zickgraf, F-J. \& Humphreys, R. M. 1991, \aj, 102, 113

\bibitem[Zickgraf, Szeifert \& Humphreys(1996)]{Z96}Zickgraf, F-J., Szeifert, Th., \& Humphreys, R. M. 1996, \aap, 312, 419 

\end{thebibliography}
\end{document}